\documentclass[traditabstract]{aa}
\usepackage{graphicx}
\usepackage{txfonts}
\usepackage{natbib}
\bibpunct{(}{)}{;}{a}{}{,}

\def\mrk509{{Mrk~509}}

\def\arcsec{\ifmmode {''} \else ${''}$\fi}
\def\arcmin{\ifmmode {'} \else ${'}$\fi}
\def\deg{\ifmmode {^\circ} \else ${^\circ}$\fi}
\def\cc{\ifmmode {\rm cm}^{-3} \else cm$^{-3}$\fi}
\def\cl{\ifmmode {\rm cm}^{-2} \else cm$^{-2}$\fi}
\def\pcm2{\ifmmode {\rm cm}^{-2} \else cm$^{-2}$\fi}
\def\micron{\ifmmode \mu{\rm m} \else $\mu$m\fi}
\def\kms{\ifmmode {\rm km\,s}^{-1} \else km\,s$^{-1}$\fi}
\def\kmps{\ifmmode {\rm km\,s}^{-1} \else km\,s$^{-1}$\fi}
\def\Hubble{\ifmmode {\rm km\,s}^{-1}\,{\rm Mpc}^{-1}
        \else km\,s$^{-1}$\,Mpc$^{-1}$\fi}
\def\ergsec{\ifmmode {\rm ergs\;s}^{-1} \else ergs s$^{-1}$\fi}
\def\ergcms{\ifmmode {\rm ergs\,cm}^{-2}\,{\rm s}^{-1}
          \else ergs\,cm$^{-2}$\,s$^{-1}$\fi}
\def\ergcmsA{\ifmmode {\rm ergs\,cm}^{-2}\,{\rm s}^{-1}\,{\rm \AA}^{-1}
          \else ergs\,cm$^{-2}$\,s$^{-1}$\,\AA$^{-1}$\fi}
\def\ergcmsHz{\ifmmode {\rm ergs\,cm}^{-2}\,{\rm s}^{-2}\,{\rm Hz}^{-1}
          \else ergs\,cm$^{-2}$\,s$^{-1}$\,Hz$^{-1}$\fi}
\def\Msun{\ifmmode M_{\odot} \else $M_{\odot}$\fi}
\def\Lsun{\ifmmode L_{\odot} \else $L_{\odot}$\fi}
\def\qo{\ifmmode q_{0} \else $q_{0}$\fi}
\def\Ho{\ifmmode H_{0} \else $H_{0}$\fi}
\def\hi{H {\sc i}}
\def\Halpha{H$\alpha$}

\def\lya{Ly$\alpha$}
\def\lyb{Ly$\beta$}
\def\lyg{Ly$\gamma$}

\def\cii{C\,{\sc ii}}
\def\ciii{C\,{\sc iii}}
\def\civ{C\,{\sc iv}}

\newcommand{\ovi}{O~{\sc vi}}

\newcommand{\heii}{He~{\sc ii}}

\def\nv{N~{\sc v}}
\def\niii{N~{\sc iii}}
\def\siiv{Si~{\sc iv}}
\def\siiii{Si~{\sc iii}}
\def\siii{Si~{\sc ii}}

\begin{document}

\title{ Multiwavelength campaign on \mrk509 }
\subtitle{VI. HST/COS observations of the far-ultraviolet spectrum
\thanks{
Based on observations made with the NASA/ESA Hubble Space Telescope, obtained
at the Space Telescope Science Institute, which is operated by the
Association of Universities for Research in Astronomy, Inc., under NASA
contract NAS 5-26555. These observations are associated with program \# 12022.
}
}
\author{ G. A. Kriss\inst{1,2}
	\and
          N. Arav\inst{3}
          \and
          J.S. Kaastra\inst{4,5}
           \and
           J. Ebrero \inst{4}
          \and
          C. Pinto\inst{4}
	\and
	  B. Borguet\inst{3}
	\and
	  D. Edmonds\inst{3}
          \and
          E. Costantini\inst{4}
	\and
          K. C. Steenbrugge\inst{6,7}
          \and
	R.G. Detmers\inst{4,5}
          \and
          E. Behar\inst{8}
          \and
          S. Bianchi\inst{9}
          \and
          A. J. Blustin\inst{10}
          \and
          G. Branduardi-Raymont\inst{11}
          \and
          M. Cappi\inst{12}
          \and
          M. Mehdipour\inst{11}
          \and
          P. Petrucci\inst{13}
          \and
          G. Ponti\inst{14}
}

\institute{
	Space Telescope Science Institute,
        3700 San Martin Drive, Baltimore, MD, 21218, USA \email{gak@stsci.edu}\label{inst1}
\and
	Department of Physics \& Astronomy, The Johns Hopkins University,
        Baltimore, MD, 21218, USA\label{inst2}
\and
    Department of Physics, Virginia Tech, Blacksburg, VA 24061, USA\label{inst3}
\and
    SRON Netherlands Institute for Space Research,
    Sorbonnelaan 2, 3584 CA Utrecht, The Netherlands\label{inst4}
    \and
    Astronomical Institute, University of Utrecht, Postbus 80000, 3508
TA Utrecht, The Netherlands\label{inst5}
    \and
    Instituto de Astronom\'ia, Universidad Cat\'olica del Norte, Avenida
Angamos 0610, Casilla 1280, Antofagasta, Chile\label{inst6}
    \and
    Department of Physics, University of Oxford, Keble Road,
    Oxford OX1 3RH, UK\label{inst7}
    \and
    Department of Physics, Technion, Haifa 32000, Israel\label{inst8}
    \and
    Dipartimento di Fisica, Universita degli Studi Roma Tre, via della
Vasca Navale 84, I-00146 Roma, Italy\label{inst9}
    \and
    Institute of Astronomy, University of Cambridge, Madingley Road,
    Cambridge CB3 0HA, UK\label{inst10}
    \and
    Mullard Space Science Laboratory, University College London,
Holmbury St. Mary, Dorking, Surrey, RH5 6NT, UK\label{inst11}
    \and
    INAF-IASF Bologna, Via Gobetti 101, I-40129 Bologna, Italy\label{inst12}
    \and
    UJF-Grenoble 1 / CNRS-INSU, Institut de Plan\'etologie et d'Astrophysique
    de Grenoble (IPAG) UMR 5274, Grenoble, F-38041, France\label{inst13}
    \and
    School of Physics and Astronomy, University of Southampton,
Highfield, Southampton SO17 1BJ, UK\label{inst14}
}

\date{
	Accepted 26 May 2011
}

\abstract{
We present medium-resolution ($\lambda / \Delta\lambda \sim 20,000$)
ultraviolet spectra covering the 1155--1760 \AA\ spectral range of
the Seyfert 1 galaxy \mrk509 obtained using the Cosmic Origins Spectrograph
(COS) on the Hubble Space Telescope (HST).
Our observations were obtained simultaneously with a Low Energy Transmission
Grating Spectrometer observation using the {\it Chandra} X-ray Observatory, and
they are part of a multiwavelength campaign in September through December 2009
which also included observations with {\it XMM-Newton}, Swift, and INTEGRAL.
Our spectra are the highest signal-to-noise observations to date of the
intrinsic absorption components seen in numerous prior ultraviolet
observations.
To take advantage of the high signal-to-noise ratio, we describe special
calibrations for wavelength, flat-field and line-spread function corrections 
that we applied to the COS data.
We detect additional complexity in the absorption troughs
compared to prior observations made with the Space Telescope Imaging
Spectrograph (STIS) on HST.
We attribute the UV absorption to a variety of sources in \mrk509,
including an outflow from the active nucleus, the interstellar medium and
halo of the host galaxy, and possible infalling clouds or stripped gaseous
material from a merger that are illuminated by the ionizing radiation of
the active nucleus.
Variability between the STIS and COS observation of the most blue-shifted
component (\#1) allows us to set an upper limit on its distance of $< 250$ pc.
Similarly, variability of component 6 between FUSE observations limits its
distance to  $<$ 1.5 kpc.
The absorption lines in all components only partially cover the emission
from the active nucleus with covering fractions that are lower than those
seen in the prior STIS observations and are comparable to those seen in spectra
from the Far Ultraviolet Spectroscopic Explorer (FUSE).
Given the larger apertures of COS and FUSE compared to STIS, we favor
scattered light from an extended region near the active nucleus as the
explanation for the partial covering.
As observed in prior X-ray and UV spectra, the UV absorption has velocities
comparable to the X-ray absorption, but the bulk of the ultraviolet
absorption is in a lower ionization state with lower total column density
than the gas responsible for the X-ray absorption.
We conclude that the outflow from the active nucleus is a multiphase wind.
}


\maketitle
%

\section{Introduction}

Outflows from galaxies powered by an active galactic nucleus (AGN) may play an
important role in the chemical enrichment of the intergalactic medium (IGM).
The most distant AGN known, quasars at redshifts of 6 or more, have
abundances at solar or super-solar levels
\citep{Hamann93, Hamann99, Pentericci02, Barth03, Dietrich03, Freudling03, Juarez09}.
Although the chemical enrichment of the host likely predates the presence of
the active nucleus \citep{Simon10}, winds powered by the AGN could return this
metal-enriched material to the IGM
\citep{Furlanetto01, Cavaliere02, Germain09, Hopkins10, Barai11}.
Their presence and feedback may also have a significant impact on the
evolution of their host galaxies
\citep{Silk98, Scannapieco04, Granato04, DiMatteo05, Hopkins08, Somerville08}.
While some fraction of these outflows in low-luminosity AGN may not escape
their host galaxy, at least as measured in the local universe
\citep{Das05, Ruiz05, Das07},
the impact of the outflow on lower-density portions of the host interstellar
medium (ISM) could provide a significant transport mechanism for the enrichment
of the surrounding environment \citep{Hopkins10}.
From constraints provided by the X-ray background, such lower luminosity AGN
may dominate the population of active galaxies in the early universe
\citep{Treister09, Treister10}.

Nearby AGN provide local analogs that can help us to understand the mechanics,
energetics, and chemical enrichment patterns that may play a significant
role in cosmic evolution at high redshift.
More than half of low-redshift AGN exhibit blue-shifted UV or X-ray
absorption features indicative of outflowing gas
\citep{Crenshaw03, Dunn07, Cappi09, Tombesi10}.
Understanding the geometry and the location of the outflow relative to the
active nucleus is a key to making an accurate assessment of the total mass
and the kinetic luminosity of the outflow.
Distance determinations are particularly difficult.
Using density-sensitive absorption lines to establish the gas density,
in combination with photoionization models that reproduce the observed
relative column densities can provide precise distance estimates.
These measures have ranged from tens of parsecs
in NGC 3783 \citep{Gabel05b} and NGC~4151
(\citealt{Kraemer06}, if one uses the correct critical density for \ciii*),
and up to kiloparsec scales in some quasars and AGN
\citep{Hamann01, Scott04a, Edmonds11}.
UV and X-ray observations of Mrk~279 \citep{Scott04a, Arav07, Costantini07}
measured absolute abundances in the outflow of a local AGN for the first time,
and they show evidence for enhanced CNO abundances that suggest contributions
from massive stars and AGB stellar winds.

To improve upon these prior studies, we have conducted a multiwavelength
campaign of coordinated X-ray, UV, and optical observations of the nearby
luminous Seyfert 1 galaxy \mrk509 (z=0.034397; \citet{Huchra93}).
A complete overview of the campaign is given by \citet{Kaastra11a}.
\mrk509 is an ideal object for study due to its high flux, moderate luminosity
that rivals that of QSOs \citep{Kopylov74}, and deep, well structured absorption
troughs.
The blue-shifted absorption indicative of a nuclear outflow has been known
since the earliest UV spectral observations \citep{Wu80, York84}.
The outflow is also apparent in the blueshifted
{[O\,{\sc iii}]} emission detected across the face of \mrk509 by \citet{Phillips83}.
More recently \citet{Kriss00} and \citet{Kraemer03} observed
\mrk509 at high spectral resolution in the UV with the
Far Ultraviolet Spectroscopic Explorer (FUSE)
and the Space Telescope Imaging Spectrograph (STIS)
on the Hubble Space Telescope (HST), respectively.
The STIS observations were simultaneous with earlier {\it Chandra} X-ray
grating observations \citep{Yaqoob03} that also detect blue-shifted X-ray
absorption lines.
These prior UV observations show the presence of more than seven distinct
kinematic components, none of which could be unambiguously associated
with the absorption detected in the X-ray, although both the X-ray and UV
absorption do overlap in velocity.

Our newer observations using the Cosmic Origins Spectrograph (COS) on HST
achieve higher signal-to-noise ratios (S/N)
and provide insight into long-term changes in the absorbing gas.
Our COS spectra provide a higher-resolution view of the velocity structure in
the outflowing gas, and sample lower ionization ionic species than are
present in the X-ray spectra. At high spectral resolution and high S/N,
we can use velocity-resolved spectral coverage of the \nv\ and \civ\ doublets to
solve for the covering fractions and column densities in the outflowing gas
\citep{Arav07}.
Using our COS observations of \lya\ together with the archival FUSE observations
of the higher-order Lyman lines, we can provide an absolute reference for
abundances relative to hydrogen, which cannot be measured directly in the
X-ray spectra. These higher S/N spectra combined with the higher S/N
X-ray spectra permit a more detailed examination of the links between
the UV and the X-ray absorbing gas in the outflow.

This paper describes the COS observations that are part of our multiwavelength
campaign on \mrk509, and presents an initial interpretation of the results.
In \S2 we describe our COS observations and our data reduction methods,
including enhanced calibrations for the COS data and corrections for the
line-spread function. In \S3 we describe our analysis of the UV spectra and
compare them to prior STIS and FUSE observations of \mrk509.
Finally, in \S4 we discuss the physical implications of our observations
for the geometry and location of the absorbing gas in \mrk509 and its
physical conditions. We end with a summary of our conclusions in \S5.

\section{Observations and Data Reduction}

As part of our extensive coordinated campaign on \mrk509 \citep{Kaastra11a},
we observed \mrk509 using the Far-Ultraviolet Channel and the medium-resolution
gratings of the Cosmic Origins Spectrograph (COS) on the
Hubble Space Telescope (HST).
Descriptions of COS and its on-orbit performance can be found in
the COS Instrument Handbook \citep{Dixon10}.
Our two visits on 2009 December 10 and 11 were simultaneous
with the {\it Chandra} observations described by \citet{Ebrero11a}.
Using gratings G130M and G160M and observing through the
Primary Science Aperture (PSA), we covered the far-ultraviolet 
spectral range from 1155 \AA\ to 1760 \AA. The instrument configurations, times
of observation, and integration times are summarized in Table \ref{ObsTbl}.
For each grating, we used only two tilts to avoid placing gaps in spectral
ranges of interest surrounding the strong emission and absorption features
in \mrk509. This resulted in almost complete spectral coverage except for a
small gap from 1560--1590 \AA. The two tilts and two exposures at each
setting resulted in four independent placements of the grid-wire shadows and
other instrumental artifacts along the spectrum.
This enhanced our ability to flat field these high signal-to-noise ratio data.
Exposure times were weighted more heavily in favor of grating G160M
since \mrk509 is fainter across this wavelength range, and the throughput of
G160M is lower relative to G130M.
This gives more uniform signal-to-noise ratios across the full spectrum.
For G130M we obtained a total exposure time of 9470 s; for G160M, the total 
exposure time was 16452 s.

\begin{table*}
  \centering
	\caption[]{COS Observations of \mrk509}
	\label{ObsTbl}
\begin{tabular}{l c c c c}
\hline\hline       
Data Set Name & Grating/Tilt  & Date & Start Time & Exposure Time\\
              &               &      &    (GMT)   & (s)\\
\hline
lbdh01010 & G130M/1309 & 2009-12-10 & 02:48:40 & 1993 \\
lbdh01020 & G130M/1327 & 2009-12-10 & 04:07:03 & 2742 \\
lbdh01030 & G160M/1577 & 2009-12-10 & 05:42:57 & 5484 \\
lbdh01040 & G160M/1589 & 2009-12-10 & 08:54:42 & 2742 \\
lbdh02010 & G130M/1309 & 2009-12-11 & 02:46:46 & 1993 \\
lbdh02020 & G130M/1327 & 2009-12-11 & 04:05:12 & 2742 \\
lbdh02030 & G160M/1577 & 2009-12-11 & 05:41:05 & 2742 \\
lbdh02040 & G160M/1589 & 2009-12-11 & 07:16:57 & 5484 \\
\hline                  
\end{tabular}

\end{table*}

Our data were processed with the COS calibration pipeline v2.11b at STScI.
This version of the pipeline does not correct for flat-field features
or the time-dependent sensitivity of COS, and there are still some
residual anomalies in the wavelength calibration.
To produce the best quality summed data set, we used a customized series
of procedures to make these corrections as described in the following sections.

\subsection{Flat-Field Corrections}

To flat field the data and correct for instrumental artifacts, we
used one-dimensional flat-field corrections developed from COS
observations of white dwarf standards
\citep{Ake10b}.
These flat fields correspond to each detector segment (A and B) for
each grating (G130M and G160M) we used.
The signal-to-noise ratio in these flat fields (in terms of purely
Poisson-distributed noise) is comparable to that in our data set.
To reduce the impact of the added noise that would result from dividing by
these data for our flat-field correction, we first smoothed the flat-fields.
It may seem counterintuitive to consider smoothing pixel-to-pixel flats when
using them to correct for pattern noise and artifacts, but there are two
fundamental reasons why this works. First, real artifacts and pattern noise
are never sharper than the intrinsic resolution of the detector, which is
approximately six pixels. Second, drifts in the grating select mechanism, which
are corrected during an observation using the TAG-FLASH observing mode,
effectively smear the pattern noise. Since our observations were longer than
any of the individual ones used to develop the one-dimensional flats, the
smearing of the pattern noise is greater. Smoothing the flats partially
corrects for this smearing. After trying a variety of smoothing widths,
we settled on smoothing the flats with a Gaussian profile having a 1-pixel
dispersion.
To align the flats with the extracted 1D spectra from each exposure, we
cross correlated the appropriate smoothed flat with the data, applied an
appropriate integer-pixel shift, and then divided it out.

Once the data were flat-field corrected, we aligned the individual exposures
by first cross-correlating them with each other and applying the closest
integer-pixel shift.
While this process produced a tremendous improvement in the data, some
residual features remained.
We then examined each individual exposure, comparing them with each other and
with the flat field used to correct them. Strong features in the flats that
did not divide out well were then flagged in the individual exposures so that
they could be excluded when we did the final combination.
Our final spectrum for each grating is the exposure-weighted combination of
the unflagged pixels in each of the individual flat-fielded and aligned
exposures. Although our data would support a signal-to-noise ratio of over
100 per pixel in the Poisson limit, we do not do much better than 60:1 per
pixel in our final spectrum.

\subsection{Wavelength Calibration}

While the nominal wavelength calibration of COS over most of the spectral range
has errors less than 15 $\kms$, this is not adequate for our analysis.
We wish to align spectral features
(both intrinsic to \mrk509 and in intervening material) more accurately
for comparison in velocity space, and for comparison to prior observations
with FUSE and STIS. There can also be residual zero-point shifts in the
wavelength scale due to offsets from the aperture center which may not have
been fully removed in the target acquisition process. Therefore, we first
adjusted the wavelength-scale zero points for our G130M and G160M spectra
by cross correlating them with the prior STIS observation of \mrk509.
For G130M, we used the wavelength region from 1245--1290 \AA, which is
dominated by the intrinsic absorption features from \lya\ and \nv. 
For G160M, we used the region from 1590--1615 \AA, centered on the intrinsic
absorption from \civ.
The intrinsic absorption features provide numerous sharp edges that result in
a strong, narrow cross-correlation peak that enhances the precision of the
alignment.
The resulting corrections were $<$ 0.1 \AA.

For the next level of correction, we used the numerous interstellar
absorption lines in our spectra that spanned the full wavelength range.
As the sight line to \mrk509 is very complex, each interstellar line
typically has multiple components. The H~I 21-cm emission in this direction
\citep{Murphy96} has a  strong flux-weighted peak at $+3.1~\kms$ in the
local standard of rest (LSR), and a weaker, broader peak at $\sim +65~\kms$
(LSR).
The +3 $\kms$ feature is present in all detected interstellar lines, and
the +65 $\kms$ one is present in the stronger lines.
In addition, there are high-velocity features at $-249$ and $-295~\kms$
visible in the strongest lines and in the higher ionization species
like \civ.
We measured the wavelengths of all these interstellar features.
The wavelength differences of our measurements relative to vacuum wavelengths
show a systematic, nearly linear trend with wavelength across each detector
segment that amount to an offset of ~0.06 \AA\ from one end of each segment
to the other. We take this as an indication that there is a slight error in
the dispersion constant for the COS wavelength calibration.
To weight our correction by the strongest features, we used only those
corresponding to the $+3$ and $+65~\kms$ components and performed a linear
fit to the wavelength offsets for each detector segment.
We then made a linear correction to the wavelength scale, with pivot
points that have zero correction about 1260 \AA\ and 1600 \AA\ so that we
keep our zero points aligned with the STIS wavelength scale.
After applying this correction, we then find a peak-to-peak variation
among the velocity offsets of all the interstellar features of
$-10$ to $+8~\rm \kms$, with a dispersion of 4.3 $\kms$.

\begin{figure*}
  \centering
   \includegraphics[width=17cm, angle=-90,, scale=0.45, trim=0 0 255 0]{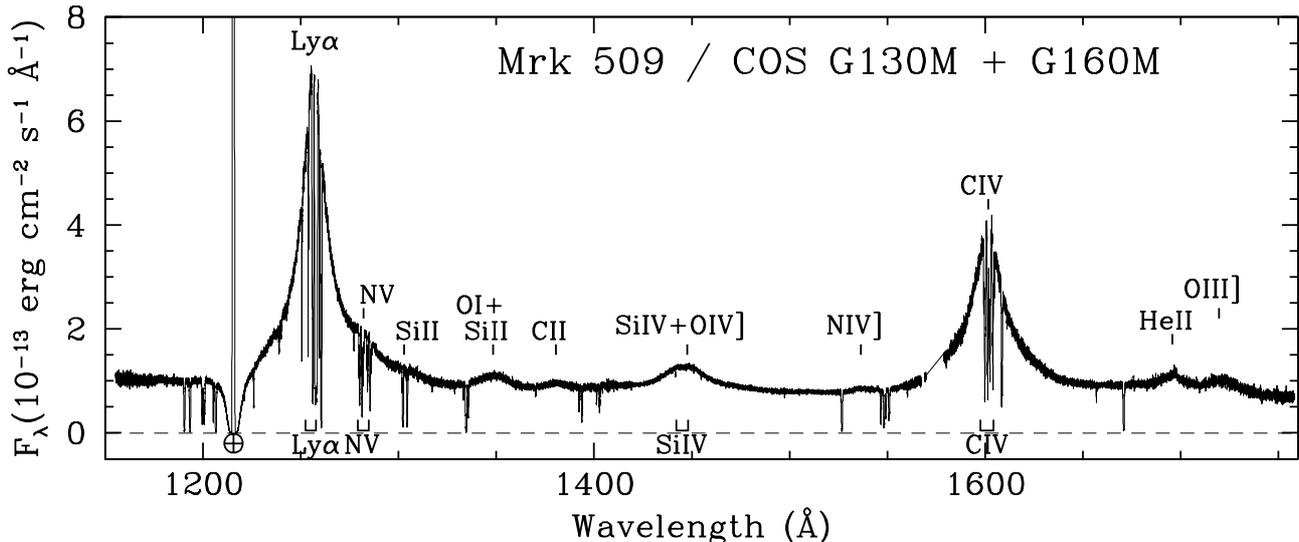}
  \caption{Calibrated and merged COS spectrum of \mrk509. Prominent emission
features are labeled above the spectrum. Regions of intrinsic absorption are
indicated below the spectrum. Geocoronal emission in the center of the galactic
\lya\ absorption trough is indicated with an Earth symbol.}
  \label{fig_cosfull}
\end{figure*}

\subsection{Flux Calibration}

The sensitivity of COS has been slowly declining since its installation
\citep{Osten10}. 
We have applied the time-dependent sensitivity corrections to our data
as given by \citet{Osten10}.
These corrections amount to an adjustment of only 2.5\% at the short wavelength end of the G130M spectrum, and range up to a correction of 5\% at the
long-wavelength end of G160M.
We estimate an absolute flux accuracy of 5\% at wavelengths above
1220 \AA, and 10\% at shorter wavelengths \citep{Massa10}.
We expect that relative fluxes are more precise.
Our repeat observations in our two visits agree to better than 1\%, and
comparison between the overlapping wavelength ranges of G130M and G160M
in our spectra show that they also agree to better than 1\%.

As an external comparison, and for cross-calibration purposes with the
other elements of our campaign, we compare our spectrum to the
Swift UVM2 data point from 2010 Dec 10.
These data are in a broad band centered at 2231 \AA, with some contamination
from broad \ciii] emission \citep{Mehdipour11}.
To compare our COS spectrum to this point requires some extrapolation.
We take the 1140--8090 \AA\ spectrum assembled by \citet{Shang05} from
FOS and KPNO 2.1-m spectra,
and normalize this spectrum to the
corrected Swift UVM2 continuum flux at 2231 \AA\ of
$5.12 \times 10^{-14}~\ergcmsA$
\citep{Mehdipour11}.
At wavelengths longer than 1220 \AA, our COS spectra agree with the
FOS spectrum normalized to the corrected Swift UVM2 flux to better than 4\%
at all wavelengths. Below 1210 \AA, our COS spectrum is brighter by up to 9\%
above the renormalized FOS spectrum. This could be a true difference in the
spectrum, but it could also be an indication of our residual uncertainties in
the flux calibration. In the latter case, we conservatively estimate that our
absolute flux below 1220 \AA\ is better than 10\%.

Figure \ref{fig_cosfull} shows our calibrated and merged COS spectra.
We have labeled the
prominent emission features in the spectrum as well as the regions
containing the intrinsic absorption features. The many galactic interstellar
absorption lines in our spectrum are not labeled here.
In an on-line appendix available only in the journal version of this paper,
Fig. A1 shows full-resolution plots of the COS spectrum of \mrk509.

\subsection{STIS Spectrum}

The STIS spectrum is a 7,600 s observation obtained on 2001-04-13 using 
the echelle E140M grating and the 0.2$\times$0.2 arc sec aperture
\citep[Proposal ID 8877, PI: Yaqoob]{Kraemer03, Yaqoob03}.
The spectrum we use is the one-dimensional extracted echelle spectrum
direct from the MAST archive with all up-to-date calibrations applied, including
corrections for scattered light in the echelle mode \citep{Valenti02} and for
the echelle blaze corrections \citep{Aloisi07}.
The STIS spectrum of \mrk509 has a resolution about twice that of 
the COS spectra, R=46,000. This spectrum is 
particularly useful for assessing the true 
resolution of the COS spectrum, and the validity of deconvolutions that 
repair the broad wings of the COS line-spread function.

\begin{figure}
  \centering
   \includegraphics[width=8.5cm, angle=-90, trim=0 30 0 0, scale=0.85]{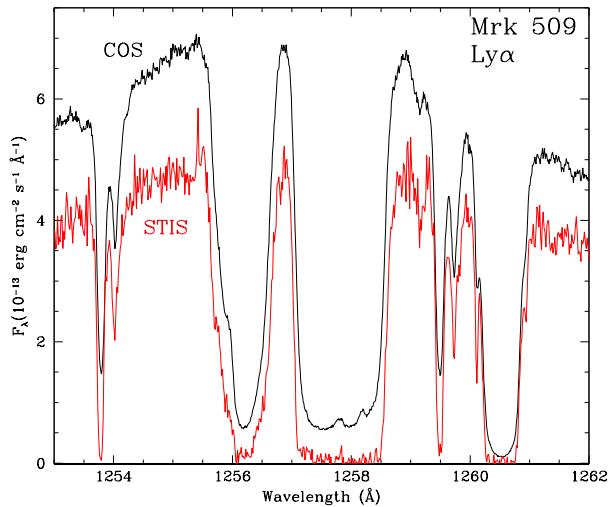}
  \caption{COS spectrum of \mrk509 in the \lya\ region (black)
compared to the STIS spectrum (red).
}
  \label{fig_cosstislya}
\end{figure}

To illustrate the dramatic improvement in the quality of our COS data
compared to the prior STIS observation,
Figure \ref{fig_cosstislya} shows a portion of both
spectra in the region centered around the \lya\ absorption feature
intrinsic to \mrk509. One sees that \mrk509 was $\sim$60\% brighter
during our COS observation. In addition, the higher throughput of COS also
improves our resulting signal-to-noise ratio by a factor of $\sim$5.
However, one can see that there are still instrumental features that we
must correct before doing a detailed analysis of the COS data.
Deep, saturated absorption features in the STIS spectrum, especially the
interstellar \siii\ line at 1260.5 \AA, do not have square, black troughs in the
COS spectrum. This is due to the broad wings on the line-spread function (LSF)
\citep{Ghavamian09} that redistribute light from the bright continuum into
cores of the absorption lines.
To extract the maximum information from our spectrum, we must correct for
these broad wings on the LSF.

\subsection{Deconvolving the COS Spectrum}

The broad wings on the COS LSF are caused by mid-frequency
polishing errors on the HST primary \& secondary mirrors \citep{Ghavamian09}.
Additional scattering by the HST mirror system and internal to COS makes these
wings more extensive than originally thought \citep{Kriss11}.
These errors must be removed in order to obtain accurate measurements 
of the depths of the absorption lines in the COS spectrum. On scales 
less than about 50 $\kms$, significant light can leak into the absorption 
line troughs from adjacent continuum regions. As a result, interstellar
absorption lines expected to be saturated are not black.
This is quite obvious when one compares the interstellar lines in the COS 
spectrum in Fig. \ref{fig_cosstislya} to the prior spectrum obtained with STIS.

\begin{figure}
  \centering
   \includegraphics[width=8.5cm, angle=-90, trim=0 30 0 0, scale=0.85]{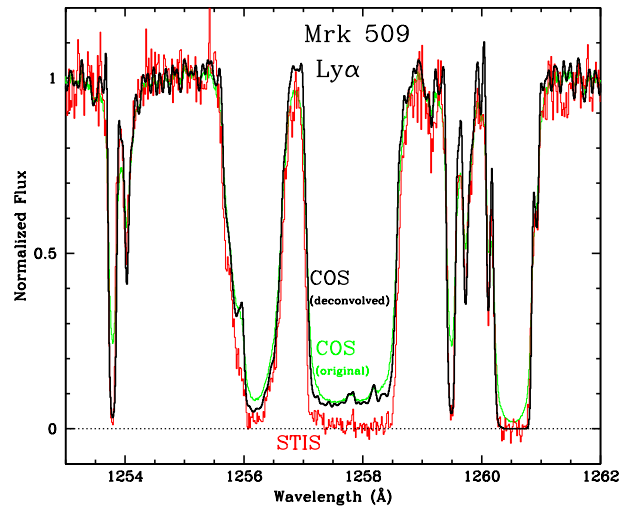}
  \caption{Comparison of the deconvolved COS spectrum in the spectral region
surrounding the \lya\ absorption trough (black) to the original
COS spectrum (green) and the STIS spectrum (red). 
}
  \label{fig_cosdecon}
\end{figure}

The high signal-to-noise ratio of our data permits us to do a fairly 
good job of deconvolving the line-spread function from our data. 
We use the Lucy-Richardson algorithm to deconvolve the data as implemented
in the STSDAS routine ``lucy" in the analysis.restore package.
We deconvolve the spectrum in 50 \AA\ intervals using the updated versions of
the wavelength-dependent LSFs from the COS web site at STScI
\citep{Kriss11}.
The deconvolution converges well after $\sim$20 iterations.
A good illustration of the success of our deconvolution is provided by
direct comparisons to the higher-resolution STIS spectrum of \mrk509.
Figure \ref{fig_cosdecon} shows the normalized version of the original COS
spectrum in the \lya\ region, the deconvolved spectrum,
and the STIS spectrum, all plotted together.
The deconvolution deepens the narrow interstellar absorption features so
that they match the higher-resolution STIS spectra well. Broader, saturated
interstellar lines like \siii\ $\lambda 1260.5$ have square, black bottoms.
Overall, the deconvolved, high S/N COS spectrum looks like an
excellent fit to the STIS spectrum.
One disadvantage to the deconvolution, however, is that it amplifies noise in
the continuum.
This can obscure weak absorption features.
Therefore, in our analysis we will adopt the approach of using the original
spectrum to identify significant features, but use the deconvolved spectrum
to make accurate measurements of the strength of known features so that
the effects of the LSF are fully corrected.

\subsection{FUSE Spectra}

FUSE \citep{Moos00} observed \mrk509 on two different occasions during the
mission.
The first FUSE spectrum, obtained in two closely spaced observations on 
1999-11-02 (19,355 s) and 1999-11-06 (32,457 s), were analyzed and
discussed by \citet{Kriss00}.
The second FUSE visit (Observation ID P1080601) was
a 62,100 s observation obtained on 2000-09-05.
We retrieved these data from the MAST archive and reprocessed them with a
customized version of the FUSE pipeline, using only the night portions
of the observation in order to minimize airglow and background.
Our version of the processing also lowers the background noise by
screening out more of the lowest pulse height channels.
We cross correlate the overlapping wavelength sections of each channel
to align the wavelength scales, and renormalize all the fluxes to match
channel LiF1A, which was aligned with the fine guidance sensor for both observations.
We correct for the ``worm" in segment LiF2A by taking a ratio to the matching
portion of the spectrum in LiF1B, fitting a spline, and applying that correction
to LiF2A. We then combine all channels in overlapping wavelength
regions weighted by the S/N of the individual pixels.
Although \mrk509 was fainter in 2000 compared to the first observation,
channel alignment was better, and the overall S/N is higher.
\mrk509 was fainter in both FUSE observations than in
our current COS observation by 34\% to 55\%.

\section{Data Analysis}

\subsection{Continuum and Emission Lines}

Our first step in analyzing the absorption features in \mrk509 is to fit the
line and continuum emission.
We use the specfit task in the contrib package of IRAF for our fits
\citep{Kriss94}.
For the continuum, we use a reddened power law 
($F_\lambda = F_{1000}~ (\lambda / 1000 \AA)^{-\alpha}$)
with extinction of E(B-V) = 0.057 (as obtained from NED via the prescription of
\citealt{Schlegel98})
and variation with wavelength as specified by \citet{CCM}
with a ratio of selective to total extinction of $\rm R_V = 3.1$.
Our best-fit power law has $\rm F_\lambda = 2.48 \times 10^{-13}~
(\lambda / 1000 \AA)^{-1.74}~\ergcmsA$.
A summary of the power law parameters and observed continuum fluxes
at 1175 \AA\ for the COS observation and the prior STIS and FUSE observations
is given in Table \ref{PLTbl}.

\begin{table*}
\centering
	\caption[]{Continuum Fits to Observations of \mrk509}
	\label{PLTbl}
\begin{tabular}{l c c c c}
\hline\hline       
Observatory & Date & $\rm F_{1000}$  & $\alpha$ & F(1175 \AA)\\
            &      &($10^{-13}~\ergcmsA$) &          & ($10^{-13}~\ergcmsA$)\\
\hline
FUSE     & 1999-11-06 & 1.58 & 0.41 & 0.607 \\
FUSE     & 2000-09-05 & 0.88 & 1.16 & 0.416 \\
HST/STIS & 2001-04-13 & 1.44 & 1.74 & 0.538 \\
HST/COS  & 2009-12-10 & 2.48 & 1.74 & 0.918 \\
\hline                  
\end{tabular}

\end{table*}

\begin{figure}
  \centering
\begin{tabular}{ c }
   \includegraphics[width=8.5cm, angle=-90, trim=0 40 0 0, scale=0.85]{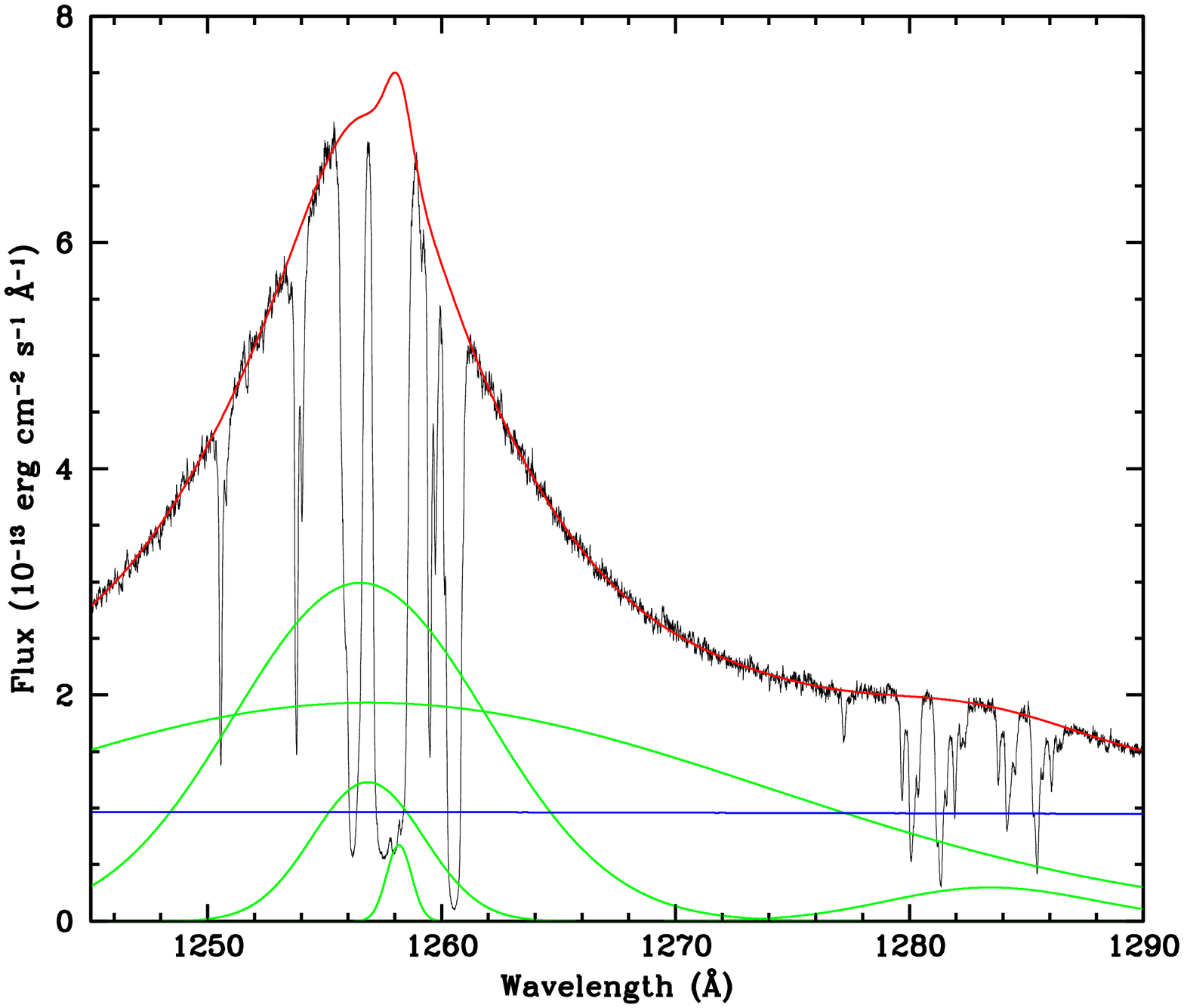}\\
   \includegraphics[width=8.5cm, angle=-90, trim=0 40 0 0, scale=0.85]{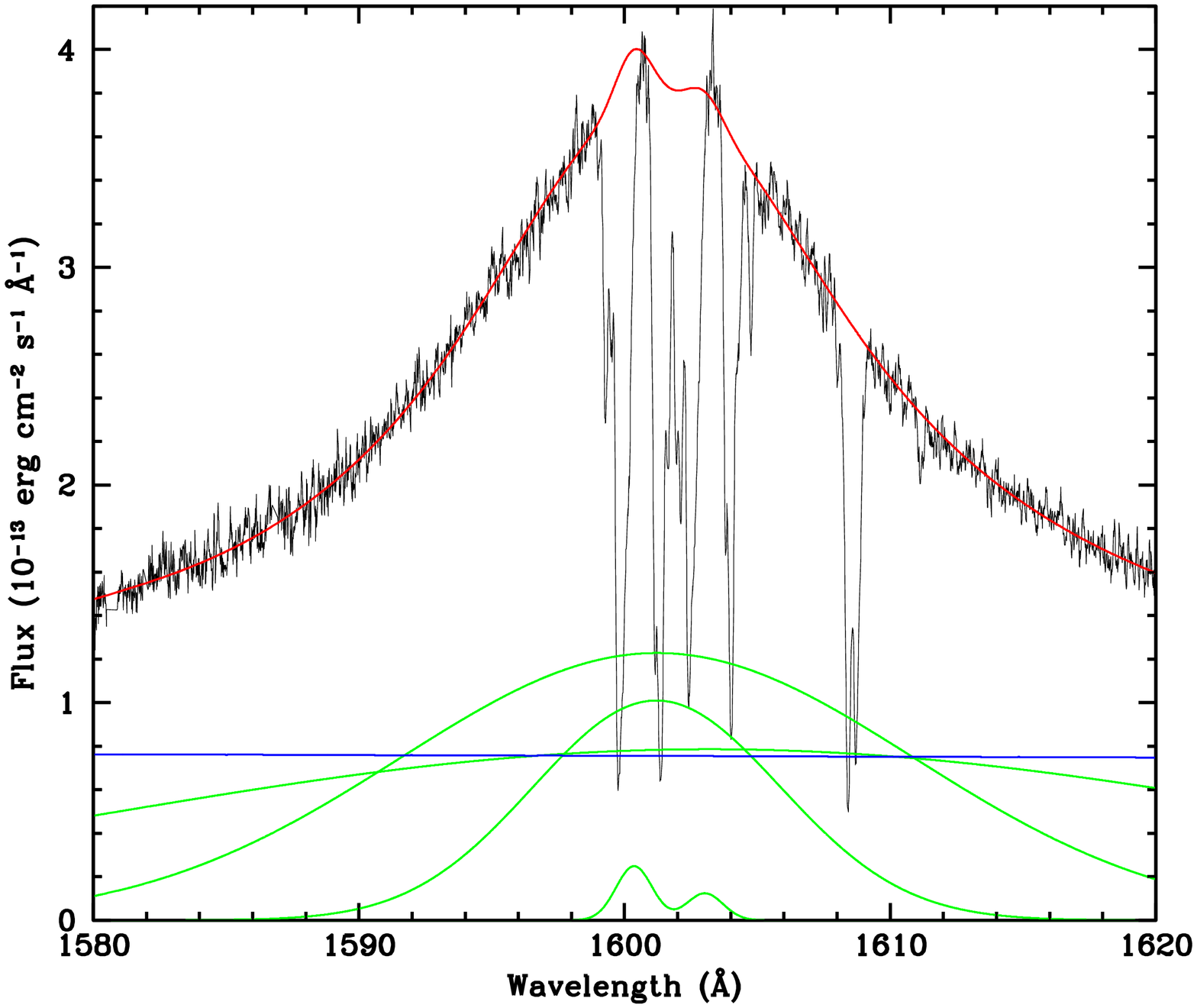}
\end{tabular}
  \caption{Fit to the broad lines and continuum of \mrk509 (red line) is
plotted on the observed spectrum (black).
Green lines show the individual emission-line components in the fit.
The blue line is the continuum.
The first panel shows the \lya\ and \nv\ region.
The second panel shows the \civ\ region.
}
  \label{fig_lyafit}
\end{figure}

For the strongest emission lines, our model requires four Gaussian
components to give a good fit. For \ovi, \lya, and \civ, we use a
broad base, two moderately broad components that comprise most of the line
profile, and a narrow-line component of full-width half-maximum (FWHM) of
300 $\kms$ fixed to match the width of the narrow {[O\,{\sc iii}]} emission
in the visible spectrum \citep{Phillips83}.
For the \ovi\ and \civ\ doublets, we include a component for each line of the
doublet with the relative wavelengths fixed at the ratio of the laboratory
values, and an optically thin, 2:1 ratio for the blue to the red flux component
of the doublet. Weaker lines in the spectrum (e.g., \ciii, \nv, \siiv)
require only a moderately broad component. The only other visible narrow
line in the far-UV spectral range is a \heii\ $\lambda 1640$ line.
The panels in Fig. \ref{fig_lyafit} show the full emission model overplotted on
the COS spectrum of \mrk509 and the individual emission components for the
regions surrounding the \lya, \nv, and \civ\ emission lines.
Tables \ref{COS_elines}, \ref{STIS_elines}, \ref{1999FUSE_elines},
and \ref{2000FUSE_elines} give the best fitting parameters for the fitted
emission lines in the COS, STIS, and FUSE spectra.
We note that the narrow-line components are weak, and they are poorly
constrained. They do improve the fit, but since they have such a small
contribution to the unabsorbed spectrum, they ultimately have little impact
on the properties derived for the absorption lines.

\subsection{Foreground absorption features}

Given our description of the emission lines and continuum in our spectrum of
\mrk509, we now identify and fit all significant absorption features in the
observed spectrum.
Most of these features are foreground absorption due to our own interstellar
medium, and they will be discussed in detail by Pinto (in preps.).
We fit the galactic and intergalactic absorption features using a Gaussian
profile in optical depth. Many of the strongest interstellar lines are
blends of features at several different velocities.
Up to seven individual components are detected in some ISM lines.
In cases where we cannot
definitively separate these components, we fix their relative velocities at the
mean of the velocities observed in cleanly resolved components of other
interstellar lines.
Our measured velocities and equivalent widths for the identified features are
given in Table \ref{tbl-ism}.
The tabulated error bars are purely statistical. For the velocities,
at a resolution of 15 \kmps\ and S/N $> 50$, the statistical errors are
expected to be $< 1$ \kmps.  In such cases, we have rounded the values up to 1.

We also find three foreground absorption features that we identify as \lya\ 
absorption in the intergalactic medium as shown in Table \ref{tbl-igm}.
One of these features was previously identified by \citet{Penton00}.
The second feature at 1221.08 \AA\ is similar to this, but much weaker.
The third feature at 1239.03 \AA\ is badly blended with foreground galactic
\nv\ $\lambda1238$. Given the observed equivalent width of the red component
of \nv\ $\lambda 1242$, this blend is too strong to be simply foreground \nv\ 
absorption. We attribute the excess (and the slight velocity offset) to an
intergalactic \lya\ absorber.
We observe no other absorption lines in our spectrum associated with these
foreground intergalactic systems.

\setcounter{table}{6}
\begin{table}
	\caption[]{Interstellar Absorption Features\\
in the COS Spectrum of \mrk509\label{tbl-ism}}
\begin{tabular}{l c c c}
\hline\hline       
Feature & $\rm \lambda_0$ & EW & $\rm V_{LSR}$ \\
        & ($\rm \AA$)  & ($\rm m\AA$)  & ($\rm km~s^{-1}$) \\
\hline
C~{\sc i}        & 1157.91 & $ 15 \pm   3$ & $    -9 \pm   1$ \\
S~{\sc iii}      & 1190.21 & $150 \pm  18$ & $    +2 \pm   1$ \\
Si~{\sc ii}      & 1190.42 & $ 75 \pm   7$ & $   -48 \pm   2$ \\
Si~{\sc ii}      & 1190.42 & $306 \pm  13$ & $    +7 \pm   1$ \\
Si~{\sc ii}      & 1190.42 & $360 \pm  34$ & $   +65 \pm   1$ \\
Si~{\sc ii}      & 1190.42 & $ 11 \pm   2$ & $  +143 \pm   1$ \\
C~{\sc i}        & 1192.22 & $ 10 \pm   2$ & $    -6 \pm   1$ \\
Si~{\sc ii}      & 1193.29 & $ 56 \pm   4$ & $   -52 \pm   1$ \\
Si~{\sc ii}      & 1193.29 & $329 \pm  18$ & $   +11 \pm   1$ \\
Si~{\sc ii}      & 1193.29 & $345 \pm  28$ & $   +72 \pm   1$ \\
Si~{\sc ii}      & 1193.29 & $ 12 \pm   2$ & $  +132 \pm   1$ \\
Mn~{\sc ii}      & 1197.18 & $ 24 \pm   3$ & $    +5 \pm   1$ \\
Mn~{\sc ii}      & 1199.39 & $ 45 \pm   6$ & $    +3 \pm   1$ \\
Mn~{\sc ii}      & 1201.12 & $ 14 \pm   2$ & $    +4 \pm   1$ \\
N~{\sc i}        & 1199.55 & $193 \pm   5$ & $    +9 \pm   1$ \\
N~{\sc i}        & 1199.55 & $171 \pm   4$ & $   +74 \pm   1$ \\
N~{\sc i}        & 1200.22 & $173 \pm   5$ & $    +8 \pm   1$ \\
N~{\sc i}        & 1200.22 & $149 \pm   3$ & $   +70 \pm   1$ \\
N~{\sc i}        & 1200.71 & $175 \pm   5$ & $    +6 \pm   1$ \\
N~{\sc i}        & 1200.71 & $115 \pm   3$ & $   +69 \pm   1$ \\
Si~{\sc iii}     & 1206.50 & $202 \pm   4$ & $  -289 \pm   1$ \\
Si~{\sc iii}     & 1206.50 & $ 59 \pm   3$ & $  -227 \pm   1$ \\
Si~{\sc iii}     & 1206.50 & $ 11 \pm   1$ & $   -63 \pm   1$ \\
Si~{\sc iii}     & 1206.50 & $417 \pm  37$ & $    +6 \pm   1$ \\
Si~{\sc iii}     & 1206.50 & $477 \pm  84$ & $   +66 \pm   1$ \\
Si~{\sc iii}     & 1206.50 & $ 26 \pm   2$ & $  +140 \pm   1$ \\
N~{\sc v}        & 1238.82 & $ 11 \pm   2$ & $  -292 \pm   1$ \\
N~{\sc v}        & 1238.82 & $  7 \pm   2$ & $  -236 \pm   1$ \\
N~{\sc v}        & 1238.82 & $ 37 \pm   2$ & $   +11 \pm   1$ \\
N~{\sc v}        & 1238.82 & $ 10 \pm   2$ & $   +69 \pm   1$ \\
N~{\sc v}        & 1242.80 & $  6 \pm   2$ & $  -294 \pm   1$ \\
N~{\sc v}        & 1242.80 & $  4 \pm   2$ & $  -238 \pm   1$ \\
N~{\sc v}        & 1242.80 & $ 19 \pm   2$ & $   +10 \pm   1$ \\
N~{\sc v}        & 1242.80 & $  5 \pm   2$ & $   +68 \pm   1$ \\
Mg~{\sc ii}      & 1239.93 & $ 27 \pm   2$ & $   +10 \pm   1$ \\
Mg~{\sc ii}      & 1240.39 & $ 16 \pm   2$ & $    +8 \pm   1$ \\
S~{\sc ii}       & 1250.58 & $117 \pm   3$ & $    +4 \pm   1$ \\
S~{\sc ii}       & 1250.58 & $ 34 \pm   2$ & $   +60 \pm   1$ \\
S~{\sc ii}       & 1253.81 & $138 \pm   3$ & $    +8 \pm   1$ \\
S~{\sc ii}       & 1253.81 & $ 72 \pm   1$ & $   +65 \pm   1$ \\
S~{\sc ii}       & 1259.52 & $151 \pm   2$ & $    +4 \pm   1$ \\
S~{\sc ii}       & 1259.52 & $ 85 \pm   1$ & $   +64 \pm   1$ \\
S~{\sc ii}       & 1259.52 & $153 \pm   1$ & $    +4 \pm   1$ \\
S~{\sc ii}       & 1259.52 & $ 86 \pm  14$ & $   +64 \pm   1$ \\
Si~{\sc ii}      & 1260.42 & $ 14 \pm   2$ & $  -291 \pm   1$ \\
Si~{\sc ii}      & 1260.42 & $  6 \pm   3$ & $   -96 \pm   1$ \\
Si~{\sc ii}      & 1260.42 & $ 51 \pm   6$ & $   -61 \pm   1$ \\
Si~{\sc ii}      & 1260.42 & $442 \pm  15$ & $   +11 \pm   1$ \\
Si~{\sc ii}      & 1260.42 & $341 \pm  11$ & $   +75 \pm   1$ \\
Si~{\sc ii}      & 1260.42 & $ 31 \pm   4$ & $  +134 \pm   1$ \\
C~{\sc i}        & 1277.25 & $ 51 \pm  32$ & $    +5 \pm   1$ \\
P~{\sc ii}       & 1301.87 & $ 41 \pm   3$ & $    +3 \pm   1$ \\
O~{\sc i}        & 1302.17 & $ 16 \pm   6$ & $  -292 \pm   1$ \\
O~{\sc i}        & 1302.17 & $  6 \pm   6$ & $   -97 \pm   1$ \\
O~{\sc i}        & 1302.17 & $  2 \pm   6$ & $   -64 \pm   1$ \\
O~{\sc i}        & 1302.17 & $307 \pm   9$ & $    +7 \pm   1$ \\
O~{\sc i}        & 1302.17 & $277 \pm  21$ & $   +74 \pm   1$ \\
O~{\sc i}        & 1302.17 & $  6 \pm   5$ & $  +131 \pm   1$ \\
Si~{\sc ii}      & 1304.37 & $ 16 \pm   6$ & $  -290 \pm   1$ \\
Si~{\sc ii}      & 1304.37 & $  6 \pm   6$ & $   -95 \pm   1$ \\
Si~{\sc ii}      & 1304.37 & $  8 \pm   6$ & $   -39 \pm  13$ \\
Si~{\sc ii}      & 1304.37 & $284 \pm  11$ & $    +9 \pm   1$ \\
Si~{\sc ii}      & 1304.37 & $251 \pm  10$ & $   +66 \pm   1$ \\
Si~{\sc ii}      & 1304.37 & $ 15 \pm   6$ & $  +133 \pm   1$ \\
C~{\sc i}        & 1328.83 & $ 27 \pm   4$ & $    -1 \pm   2$ \\
C~{\sc ii}       & 1334.53 & $ 76 \pm   4$ & $  -293 \pm   1$ \\
C~{\sc ii}       & 1334.53 & $ 23 \pm   4$ & $  -223 \pm   1$ \\
\hline                  
\end{tabular}
\end{table}
\setcounter{table}{6}
\begin{table}
	\caption[]{Interstellar Absorption Features\\
in the COS Spectrum of \mrk509 (cont.)}
\begin{tabular}{l c c c}
\hline
C~{\sc ii}       & 1334.53 & $ 65 \pm   4$ & $   -59 \pm   1$ \\
C~{\sc ii}       & 1334.53 & $496 \pm  37$ & $   +11 \pm   1$ \\
C~{\sc ii*}      & 1335.50 & $444 \pm  21$ & $  -144 \pm   1$ \\
C~{\sc ii*}      & 1335.50 & $150 \pm   4$ & $   +49 \pm   1$ \\
C~{\sc ii*}      & 1335.50 & $105 \pm   4$ & $  +105 \pm   1$ \\
Ni~{\sc ii}      & 1317.22 & $ 33 \pm   8$ & $    +3 \pm   3$ \\
Ni~{\sc ii}      & 1317.22 & $ 16 \pm   3$ & $   +64 \pm   1$ \\
Ni~{\sc ii}      & 1370.13 & $ 41 \pm   3$ & $    +4 \pm   1$ \\
Si~{\sc iv}      & 1393.76 & $ 95 \pm   4$ & $  -298 \pm   1$ \\
Si~{\sc iv}      & 1393.76 & $ 16 \pm   3$ & $  -246 \pm   1$ \\
Si~{\sc iv}      & 1393.76 & $ 19 \pm   3$ & $   -63 \pm   1$ \\
Si~{\sc iv}      & 1393.76 & $227 \pm  14$ & $    +8 \pm   2$ \\
Si~{\sc iv}      & 1393.76 & $165 \pm  18$ & $   +64 \pm   3$ \\
Si~{\sc iv}      & 1393.76 & $  7 \pm   5$ & $  +131 \pm   1$ \\
Si~{\sc iv}      & 1402.77 & $ 53 \pm   3$ & $  -297 \pm   1$ \\
Si~{\sc iv}      & 1402.77 & $ 11 \pm   3$ & $  -275 \pm   3$ \\
Si~{\sc iv}      & 1402.77 & $  7 \pm   3$ & $   -64 \pm   1$ \\
Si~{\sc iv}      & 1402.77 & $144 \pm   7$ & $    +9 \pm   1$ \\
Si~{\sc iv}      & 1402.77 & $103 \pm   8$ & $   +65 \pm   2$ \\
Si~{\sc iv}      & 1402.77 & $  2 \pm   1$ & $  +131 \pm   1$ \\
Ni~{\sc ii}      & 1454.84 & $ 48 \pm   5$ & $    -6 \pm   1$ \\
Si~{\sc iv}$^{\rm a}$  & 1393.76 & $ 89 \pm   4$ & $  -298 \pm   1$ \\
Si~{\sc iv}      & 1393.76 & $ 14 \pm   3$ & $  -252 \pm   3$ \\
Si~{\sc iv}      & 1393.76 & $ 19 \pm   3$ & $   -67 \pm   1$ \\
Si~{\sc iv}      & 1393.76 & $223 \pm   8$ & $    +4 \pm   1$ \\
Si~{\sc iv}      & 1393.76 & $160 \pm   8$ & $   +60 \pm   1$ \\
Si~{\sc iv}      & 1393.76 & $ 20 \pm   4$ & $  +126 \pm   1$ \\
Si~{\sc iv}      & 1402.77 & $ 65 \pm   2$ & $  -297 \pm   1$ \\
Si~{\sc iv}      & 1402.77 & $ 14 \pm   2$ & $  -250 \pm   3$ \\
Si~{\sc iv}      & 1402.77 & $  7 \pm   2$ & $   -62 \pm   1$ \\
Si~{\sc iv}      & 1402.77 & $157 \pm   7$ & $    +9 \pm   1$ \\
Si~{\sc iv}      & 1402.77 & $116 \pm   8$ & $   +69 \pm   1$ \\
Si~{\sc iv}      & 1402.77 & $  2 \pm   2$ & $  +131 \pm   1$ \\
Ni~{\sc ii}      & 1454.84 & $ 24 \pm   2$ & $    -2 \pm   2$ \\
Ni~{\sc ii}      & 1454.84 & $ 11 \pm   2$ & $   +56 \pm   1$ \\
Si~{\sc ii}      & 1526.71 & $ 25 \pm   2$ & $   -73 \pm   1$ \\
Si~{\sc ii}      & 1526.71 & $408 \pm  11$ & $    +0 \pm   1$ \\
Si~{\sc ii}      & 1526.71 & $340 \pm  11$ & $   +59 \pm   1$ \\
C~{\sc iv}       & 1548.19 & $245 \pm   5$ & $  -298 \pm   1$ \\
C~{\sc iv}       & 1548.19 & $ 78 \pm   5$ & $  -245 \pm   1$ \\
C~{\sc iv}       & 1548.19 & $ 19 \pm   3$ & $  -125 \pm   3$ \\
C~{\sc iv}       & 1548.19 & $ 22 \pm   4$ & $   -63 \pm   1$ \\
C~{\sc iv}       & 1548.19 & $336 \pm   8$ & $    +8 \pm   1$ \\
C~{\sc iv}       & 1548.19 & $293 \pm   9$ & $   +66 \pm   1$ \\
C~{\sc iv}       & 1548.19 & $ 18 \pm   4$ & $  +130 \pm   1$ \\
C~{\sc iv}       & 1550.77 & $170 \pm   4$ & $  -298 \pm   1$ \\
C~{\sc iv}       & 1550.77 & $ 52 \pm   4$ & $  -248 \pm   1$ \\
C~{\sc iv}       & 1550.77 & $  1 \pm  14$ & $  -269 \pm  11$ \\
C~{\sc iv}       & 1550.77 & $ 10 \pm   4$ & $   -64 \pm   1$ \\
C~{\sc iv}       & 1550.77 & $251 \pm   7$ & $    +5 \pm   1$ \\
C~{\sc iv}       & 1550.77 & $178 \pm   7$ & $   +65 \pm   1$ \\
C~{\sc iv}       & 1550.77 & $ 11 \pm   4$ & $  +129 \pm   1$ \\
C~{\sc i}        & 1560.31 & $ 45 \pm   4$ & $    -2 \pm   1$ \\
Fe~{\sc ii}      & 1608.45 & $ 22 \pm   2$ & $   -73 \pm   1$ \\
Fe~{\sc ii}      & 1608.45 & $217 \pm   5$ & $    +2 \pm   1$ \\
Fe~{\sc ii}      & 1608.45 & $204 \pm   5$ & $   +59 \pm   1$ \\
Fe~{\sc ii}      & 1608.45 & $ 19 \pm   3$ & $   +91 \pm   1$ \\
Fe~{\sc ii}      & 1611.20 & $ 36 \pm   4$ & $    +2 \pm   2$ \\
C~{\sc i}        & 1656.93 & $ 72 \pm   5$ & $    -3 \pm   1$ \\
C~{\sc i}        & 1656.93 & $ 15 \pm   3$ & $   +59 \pm   1$ \\
Al~{\sc ii}      & 1670.79 & $ 33 \pm   3$ & $   -64 \pm   8$ \\
Al~{\sc ii}      & 1670.79 & $420 \pm  30$ & $    +3 \pm   1$ \\
Al~{\sc ii}      & 1670.79 & $379 \pm  28$ & $   +56 \pm   2$ \\
Al~{\sc ii}      & 1670.79 & $ 23 \pm   5$ & $  +121 \pm   4$ \\
Ni~{\sc ii}      & 1709.60 & $ 63 \pm  12$ & $   +11 \pm   4$ \\
Ni~{\sc ii}      & 1709.60 & $ 28 \pm   6$ & $   +69 \pm   4$ \\
Ni~{\sc ii}      & 1741.55 & $ 50 \pm  10$ & $    +8 \pm   6$ \\
Ni~{\sc ii}      & 1741.55 & $ 20 \pm   9$ & $   +66 \pm   6$ \\
Ni~{\sc ii}      & 1751.91 & $ 36 \pm  13$ & $    +3 \pm   5$ \\
Ni~{\sc ii}      & 1751.91 & $ 17 \pm  10$ & $   +61 \pm   5$ \\
\hline                  
\end{tabular}
\parbox{2.25in}{\noindent
$^{\rm a}$The prior entries for Si {\sc iv} $\lambda\lambda$1393,1402 and
Ni {\sc ii} $\lambda$1454 were for features in the G130M spectrum.
These subsequent entries are for features detected in the G160M spectrum.
} 
\end{table}

\setcounter{table}{7}
\begin{table}
  \centering
	\caption[]{Intergalactic Absorption Features\\
in the COS Spectrum of \mrk509\label{tbl-igm}}
\begin{tabular}{l c c c}
\hline\hline       
Feature & $\rm \lambda_0$ & EW & $\rm v_{\sun}$ \\
        & ($\rm \AA$)  & ($\rm m\AA$)  & ($\rm km~s^{-1}$) \\
\hline
H~{\sc i}        & 1215.67 & $ 31 \pm   3$ & $ +1345 \pm   2$ \\
H~{\sc i}        & 1215.67 & $200 \pm   4$ & $ +2561 \pm   1$ \\
H~{\sc i}        & 1215.67 & $ 72 \pm   4$ & $ +5772 \pm   2$ \\
\hline                  
\end{tabular}
\end{table}

\subsection{The Intrinsic Absorption Features}

\begin{figure}
  \centering
   \includegraphics[width=8.5cm, trim=72 72 72 72]{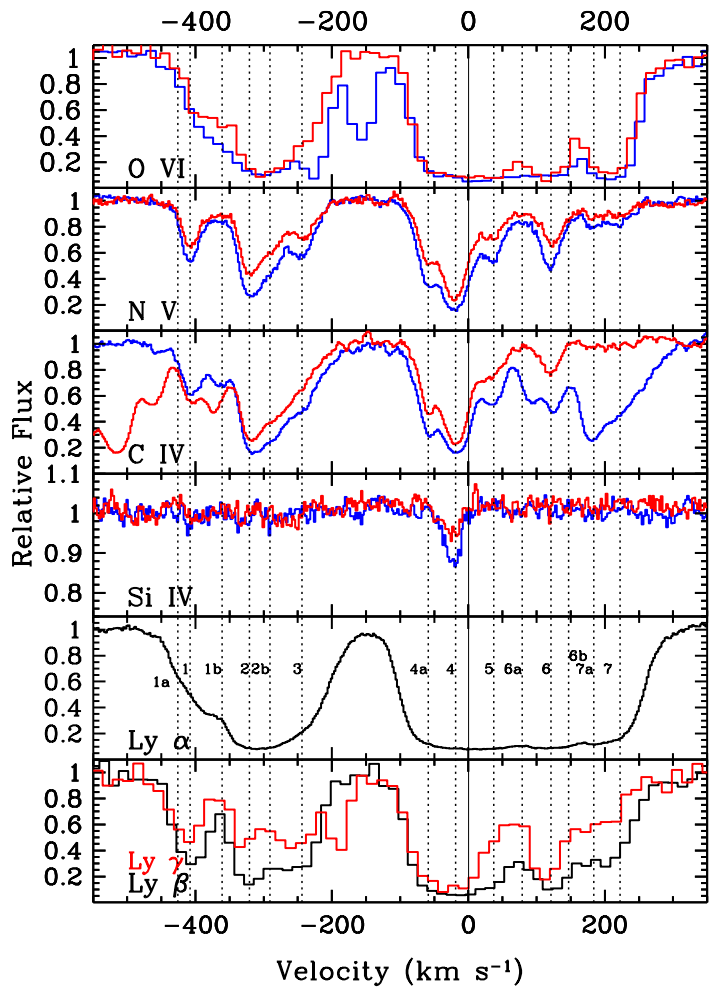}
  \caption{Intrinsic absorption features in the COS and FUSE spectra of \mrk509.
Normalized relative fluxes are plotted as a function of velocity relative to
the systemic redshift of $z=0.034397$.
The top panel shows the \ovi\ doublet from the FUSE observation in
2000 September with the red side of the doublet as a red line and the blue
side of the doublet as a blue line.
The middle four panels show the \nv, \civ, \siiv\ and \lya\ absorption troughs
from our COS observation.
The bottom panel shows the \lyg\ (red) and \lyb\ (black) absorption troughs
from the FUSE observation.
Dotted vertical lines indicate the velocities of identified absorption
components. The thin black vertical line shows $\rm v= 0$.
}
  \label{fig_cosvel}
\end{figure}

The COS spectrum of \mrk509 shows a wealth of detail in the
intrinsic absorption features.
The seven major components originally identified by \citet{Kriss00}
in the FUSE spectrum were expanded to eight by \citet{Kraemer03} based on their
STIS spectrum. Our COS spectrum shows additional subtle inflections
indicating that the absorption troughs are a quite complex
blend of perhaps even more components than we can readily resolve.
Figure \ref{fig_cosvel} shows the intrinsic absorption features detected in our
COS spectrum as a function of velocity with the components labeled.
For reference, we also include the \ovi, \lyb\ and \lyg\ regions of
the spectrum from the FUSE observation of \mrk509 obtained in 2000.
The numbering scheme follows that originated by \citet{Kriss00}, but
includes the additional component $4'$ added by \citet{Kraemer03}.
However, instead of continuing to designate
additional components with a ``prime", we use designations of ``a" or ``b"
for the additional features, with the convention that an ``a" subcomponent
is blueshifted from the main feature, and a ``b" subcomponent is redshifted.
Component $4'$ in \citet{Kraemer03} becomes 4a here.

\begin{figure}
  \centering
   \includegraphics[width=8.5cm, angle=-90, trim=0 30 0 0, scale=0.85]{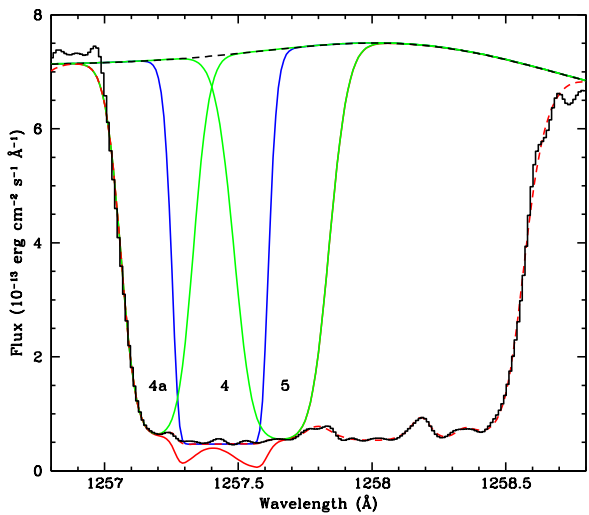}
  \caption{
Fit to a portion of the Ly$\alpha$ absorption trough in the COS spectrum
of \mrk509.
The data are shown by the solid black histogram.
The red dashed line shows the best fit model using the scattered-light/single
partial covering fraction described above for $f_c = 0.064$.
The black dashed line shows the emission model for this region.
The absorption in this model due to components 4a (shown in green), 4 (shown in blue) and 5 (shown in green) are plotted individually.
The solid red line shows the resulting model if these absorbing components
each have independent partial covering factors
$f_{c,4a} = f_{c,4} = f_{c,5} = 0.064$.
The dips in the region of overlap between each component are
prominent at wavelengths of $\sim 1257.3$ \AA\ and $1257.6$ \AA.
}
  \label{fig_pcover}
\end{figure}

To fit these features with a simple Gaussian decomposition of the profiles as
done by \citet{Kriss00} clearly has limitations built
in by the choice of the number and placement of components.
Given the complexity of the absorption troughs, a more model-independent
approach is preferred. 
Several previous studies of AGN outflows have used
absorption-line doublets to measure the optical depth and covering
fraction in discrete, velocity resolved bins (e.g.,
\citealt{Hamann97};
\citealt{Barlow97};
\citealt{Arav99};
\citealt{Ganguly99};
NGC~5548: \citealt{Arav02};
NGC~3783: \citealt{Gabel03};
Mrk~279: \citealt{Scott04a, Gabel05a}).
We will apply this method to our spectrum of \mrk509 in a subsequent
publication (Arav, in preps.).
In this paper, however, we obtain a preliminary assessment of the column
densities and covering fractions of the intrinsic absorption components by
fitting the features with Gaussian profiles in optical depth.
This enables a more straightforward comparison of our current results to the
prior FUSE and STIS observations.
Our model also allows for partial covering by each component, as well as
a scattered light component that is a wavelength-independent
fraction of the underlying emission components.
We note that in the case of a single absorbing region with a single partial 
covering fraction (denoted $f_c$) that this is observationally
indistinguishable from having a scattered light component at a level of
$(1 - f_c)$. However, in the case where there is more than one absorbing
component that overlaps in velocity with another absorbing component, there is
a significant difference. Assuming that the absorbers are statistically
independent, in the region of velocity/wavelength space that they overlap,
the transmission of light from the source will be less than for either
component individually, being the product $(1 - f_{c1})(1 - f_{c2})$.
This produces extra dips in the modeled spectrum in the regions of overlap
as illustrated in Fig. \ref{fig_pcover}.
This is not a
qualitatively accurate representation of the profiles of the absorption troughs
in our spectrum, which tend to show more smoothly varying profiles.
Characterizing the residual light at the bottoms of our absorption troughs with
a scattered component that is independent of wavelength provides a better match
to the absorption profiles that we observe.
For the \nv, \siiv, and \civ\ doublets, we link the relative wavelengths at the
ratio of their vacuum wavelengths, and we link the relative optical depths 
of the red and blue components at a value of 0.5.
Since the \lya\ absorption appears to be heavily saturated, as noted
previously by \citet{Kraemer03}, we fix the covering fraction for each of the
\lya\ components at the average of the value determined for the corresponding
components in the \civ\ and \nv\ profiles.

Our best-fit results are given in Table \ref{COS_alines}.
Note that for \siiv, as in \citet{Kraemer03}, we detect only component 4.
Also, as in \citet{Kraemer03}, no lower ionization lines (e.g.,
\siiii\ $\lambda1206$, \siii\ $\lambda1260$, or \cii\ $\lambda1335$)
are detected.
As we noted in \S3.1, the narrow emission-line components in our emission
model have little impact on the measured properties of the absorption lines.
Even if we eliminate the narrow-line emission from our fits, the measured
columns densities change by $<$ 1\%.

Using the additional components that we have identified in our COS spectra,
we have re-analyzed the archival FUSE and STIS spectra of \mrk509 in order
to make a direct comparison among the spectra.
While \citet{Kriss00} have done this for the first FUSE observation,
similar measurements for the second observation are useful since they
illustrate how the absorbers may change in response to changes in the
continuum strength.

Our results from fitting the STIS spectrum are given in Table \ref{STIS_alines}.
The column densities for the individual components agree well
(and within the errors) with those measured by \citet{Kraemer03} except for
the blended components comprising features 2 and 3, and 6 and 7.
The apportionment among the various subcomponents accounts
for these differences in our fit.
We also find a higher column density in the strongest feature, number 4,
largely because our width for that component is narrower than that
measured by \citet{Kraemer03}.

Results from fits to \lyb, \lyg\ and \ovi\ in the 1999 FUSE spectrum are in
Table \ref{1999FUSE_alines},
and results for the 2000 FUSE observation are in Table \ref{2000FUSE_alines}.
Since the \lya\ absorption trough in \mrk509 is heavily saturated,
the higher-order Lyman lines with good S/N in the FUSE spectrum are
particularly useful in deriving reliable column densities and covering factors
for \hi. However, given potential variability, these values must be used with
caution when combined with the data from our COS observations.

In carrying out our fits we found that the well defined features in \nv\ and
\civ\ in the COS spectrum did not always match well with the absorption
profiles in \lya\ and in \ovi.
One can see this in Fig. \ref{fig_cosvel}, where the \lya\ and \ovi\ profiles
vary more smoothly.  For our best-fit results in
Tables \ref{COS_alines}--\ref{2000FUSE_alines}, one can see that
some components, such as 1a, are significant only in \lya, while others, such
as 1b, are strongest in \ovi. There are also small velocity shifts for other
components when comparing \nv\ and \civ\ to \lya\ and \ovi.
This illustrates the limitations of our Gaussian decomposition, and in
\S4.2 of the Discussion we address the probable presence of multiple phases
of gas at each velocity in the intrinsic absorption troughs.

\subsection{Comparison among the COS, STIS and FUSE Spectra}

As one can see from the fits to the continua of each of the observations,
we have sampled the emission from \mrk509 over a range of about a factor of two
in overall luminosity. The FUSE observation in September 2000 represents the
faintest state ($0.42 \times 10^{-13}~\ergcmsA$ at 1175 \AA), and the COS
observation reported here is the brightest
($0.92 \times 10^{-13}~\ergcmsA$).

\begin{figure}
   \includegraphics[width=8.5cm, trim=72 72 72 72]{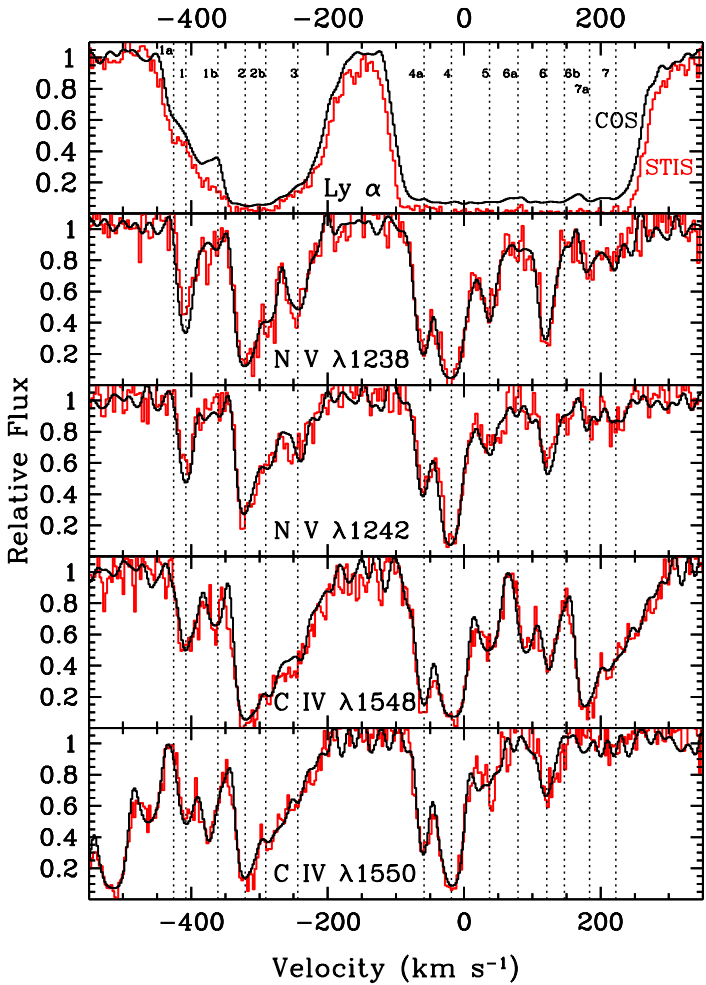}
  \caption{Comparison of spectral features in the COS (black) and
STIS (red) spectra of \mrk509.
Normalized relative fluxes are plotted as a function of velocity relative to
the systemic redshift of $z=0.034397$.
The individual panels (from top to bottom) show \lya, \nv\ $\lambda 1238$,
\nv\ $\lambda 1242$, \civ\ $\lambda 1548$, and \civ\ $\lambda 1550$.
}
  \label{fig_cosstisall}
\end{figure}

Our COS spectrum of \mrk509 captures the active nucleus in a significantly
brighter state compared to STIS. Overall, the continuum is $\sim$80\% brighter.
Despite the difference in brightness, the continuum shape does not noticeably
change between the STIS and the COS observation.
The broad emission lines are also brighter, but not by as much---only 37\%.
Despite the difference in the continuum level, there are only subtle
differences in the absorption-line troughs.
The much higher S/N in our COS spectrum reveals some interesting differences
from the original STIS spectrum.
We show this by comparing the deconvolved COS absorption-line profiles to the
STIS profiles in Fig. \ref{fig_cosstisall}.
The biggest differences are in the \lya\ absorption profile.
Significantly less absorption is present in the most-blue-shifted portions of
the trough (essentially components 1--1b). The most striking difference,
however, is the extra light that seems to fill the bottom of the \lya\ 
absorption troughs in the COS spectrum, even in the deconvolved version.
Comparable differences in the \nv\ and \civ\ troughs are not readily visible
in Fig. \ref{fig_cosstisall},
so we have carefully examined the COS data to rule out
any instrumental effects that might be causing this.

First, we can rule out instrumental scattered light, or residual effects of the
line spread function.
Since different grating tilts illuminate the detector slightly differently,
we compared the separate spectra from each grating tilt and each visit.
The extra light in the \lya\ trough is visible at both grating
tilts, 1309 and 1327, in both visits of our observations.
One can readily see in the deconvolved COS spectrum
in Fig. \ref{fig_cosdecon} that the saturated galactic \siii\ $\lambda$1260
absorption line
is fully corrected to a black trough that is a good match to the STIS
spectrum. The much deeper and wider absorption trough at galactic \lya\ at
1216 \AA\ is also black, even in the spectrum that has not been deconvolved
if one looks at regions that are more than a resolution element removed from
the geocoronal \lya\ line at the center of the trough.
(The peak of the geocoronal \lya\ emission is far brighter
than the peak of \lya\ in \mrk509, yet the flux drops quickly by over two
orders of magnitude from the peak down to the trough of the galactic \lya\ 
absorption.)

Another possibility is that extended emission in \mrk509,
in particular, extended narrow \lya\ emission might be filling in the
absorption trough. The STIS observation used the 0.2$\times$0.2 arc sec
aperture, while the COS entrance aperture is 2.5$\arcsec$ in
diameter. Extended emission that is excluded by the STIS aperture could be
entering the COS aperture. 
We examined the 2D spectra, both visually
and as summed projections showing the spatial profile, and there is no
difference in the spatial extent of the light in the \lya\ troughs
compared to the continuum.
The cross-dispersion profile of the continuum spectrum also shows the
nominal width and shape of a point source in the COS aperture.
In contrast, the spatial extent of the
geocoronal \lya\ emission, which fills the aperture, is quite
obvious. Of course it is quite bright, but even scaled down it is easily
seen to be extended. We conclude that there is no evidence for extended
line emission visible in the COS aperture.

Looking closely at the STIS spectrum, one sees that the \lya\ absorption 
is not black at all portions of the trough.
In fact, \citet{Kraemer03} required components with partial
covering to fit the intrinsic absorption lines in their spectrum (\civ\ and
\nv\ as well as \lya), as did \citet{Kriss00} in their FUSE spectrum.
If one compares the partial
covering fractions among all three data sets (FUSE, STIS, and COS), one sees
that COS and FUSE require similar covering fractions, and STIS somewhat less.
The total depth of the \ovi\ absorption troughs in \citet{Kriss00} are almost
directly comparable to the most highly saturated portions of the \lya\ trough
in the COS spectrum, with the deepest portions having $\sim92$\% coverage
(including both the scattered light components and the partial covering
factors).
Note that both the COS and FUSE spectra were obtained through large apertures
(2.5$\arcsec$ diameter and $30\arcsec \times 30\arcsec$, respectively),
suggesting that an extended emission component (or light from an extended
scattering region) is filling in the absorption troughs.
From our inspection of the COS two-dimensional images described above, since we
do not resolve this component in the cross-dispersion direction, we conclude
that such a region must be larger than the STIS aperture size, but smaller than
the nominal COS cross-dispersion spatial resolution, which is $0.5 \arcsec$
at its best at a wavelength of 1600 \AA\ \citep{Ghavamian10}.
Comparing the best-fit scattered light components from our fits to the COS and
STIS spectra, 0.05 vs. 0.01, if we assume that this light is coming from a
region of uniform surface brightness, the total extent of this region would be
$\sqrt 5$ times the size of the STIS aperture, or roughly $0.45 \arcsec$
in diameter, which is consistent with the limits in size we set from the
lack of any detectable spatial extent in the COS spectrum.

This scattering region would then have a radius of $\sim 150$ pc.
If it occupied a biconical region with a covering fraction of
$\Delta\Omega / 4 \pi = 0.25$, following \citet{Krolik86}, the required Thomson
optical depth $\tau_e$ to produce the scattered flux would be
$\tau_e = 0.05 / (\Delta\Omega / 4 \pi) = 0.20$, where 0.05 is the fraction of
uniformly scattered light in the troughs of the COS and FUSE spectra.
This gives a total column density of $3.0 \times 10^{23}~\rm cm^{-2}$ for the
hot gas.  If it uniformly fills this volume, it has a density of
$1900~\rm cm^{-3}$ and a total mass of $2.0 \times 10^8$ \Msun.
While this is larger than the reflecting region in NGC 1068, its properties
are not unreasonable for a volume filled by the outflow from the nuclear
region of \mrk509.

\begin{figure}
  \centering
   \includegraphics[width=8.5cm, trim=72 72 72 72]{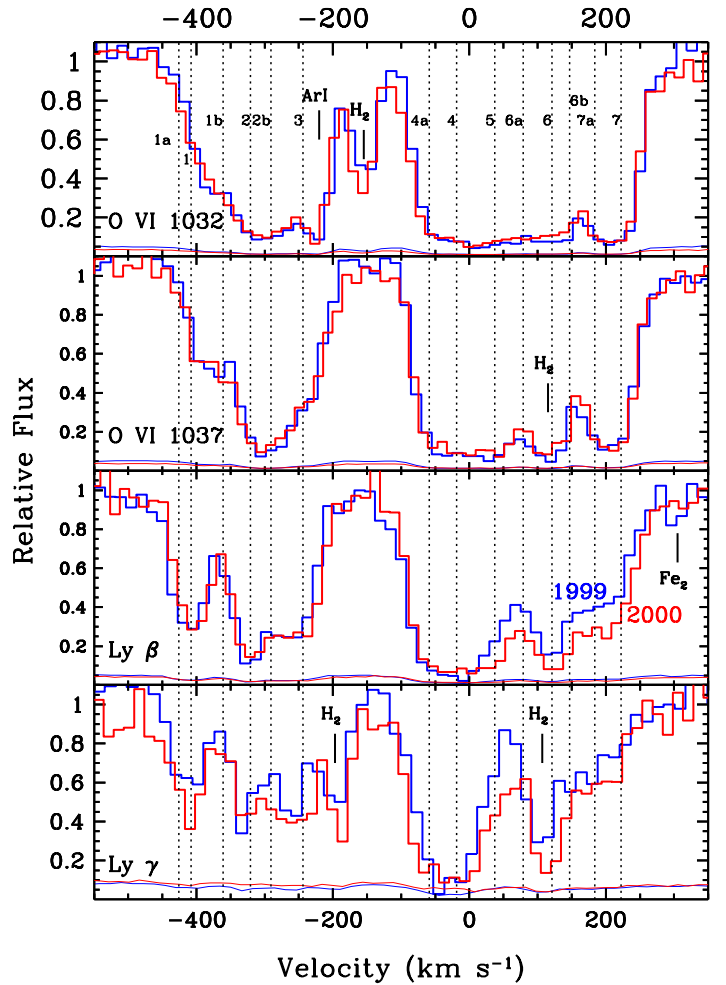}
  \caption{Comparison of spectral features in the FUSE spectra of \mrk509
from 1999 November (blue) and 2000 September (red).
Normalized relative fluxes are plotted as a function of velocity relative to
the systemic redshift of $z=0.034397$.
The individual panels (from top to bottom) show \ovi\ $\lambda 1032$,
\ovi\ $\lambda 1037$, \lyb, and \lyg.
Foreground Galactic ISM features are marked.
Error vectors for each spectrum are shown as thin lines in the same color as
the data.
}
  \label{fig_fusecomp}
\end{figure}

In addition to this scattered light component, we see differences in the depths
of several absorption components in the \lya\ and \nv\ profiles.
Examining Fig. \ref{fig_cosstisall} once again, one can see that components
1 and 1b in the COS \lya\ spectrum are not as deep as in the STIS spectrum.
In contrast, component 1 in the \nv\ absorption profile is noticeably
shallower in the STIS spectrum compared to the COS spectrum.
More quantitatively, we can compare the column densities in
Tables \ref{COS_alines} and \ref{STIS_alines}.
From the FUSE \lyg\ and \lyb\ profiles in the velocity range of components
1b, 1, and 1a we can see that this region of velocity space is optically thin,
even in the \lya\ profile. Therefore, the change in the summed column density
from these three components from $88 \times 10^{12}~\rm \pcm2$ to
$57 \times 10^{12}~\rm \pcm2$ is a decrease of 35\%.
For \nv, the column density increased from $112 \times 10^{12}~\rm \pcm2$ to
$135 \times 10^{12}~\rm \pcm2$.
If these changes are due to a change in the ionization state of the absorbing
gas, it indicates that the gas must have been in a lower ionization state in
2000, at an ionization parameter somewhat below the peak for the ionization
state of \nv.

Changes in absorption are also apparent when one compares the two FUSE
observations, obtained roughly 10 months apart.
In Fig. \ref{fig_fusecomp} we compare the absorption profiles in velocity
space for the \ovi\ doublet and for \lyb\ and \lyg.
(One must be careful in comparing these spectra since portions of each line
profile are contaminated by foreground Galactic absorption.) We have marked
the contaminating features in each panel of Fig. \ref{fig_fusecomp}.
\ovi\ is very heavily saturated at almost all velocities;
we see no significant differences visually in Fig. \ref{fig_fusecomp}.
The fits in Tables \ref{1999FUSE_alines} and \ref{2000FUSE_alines}
also show similar column densities, but, given the
level of saturation in \ovi, these are not reliable comparisons.
In comparing the \lyb\ and \lyg\ profiles, however, one can see significant
changes in the red side of the absorption troughs, especially for
components 5, 6a, 6, 6b.
There are also slight differences in the region of component 1, as we saw in
comparing the COS and STIS observations.
Doing a quantitative bin-by-bin comparison in velocity space over the
range of components 1a, 1, and 1b, and 6a, 6, and 6b, we find that changes
in the component 1 region, which primarily show up in \lyg\ are significant
at only the 1-$\sigma$ confidence level.
For components 5, 6a, 6, and 6b, however, the changes in \lyb\ and \lyg\ are
significant at $> 3\sigma$ confidence. 
Summing over all of these components for \hi\ in
Tables \ref{1999FUSE_alines} and \ref{2000FUSE_alines}, we see that the
total column of \hi\ increased from $1.16 \times 10^{15}~\rm \pcm2$ to
$1.30 \times 10^{15}~\rm \pcm2$ from 1999 November to 2000 September.
If due to a change in ionization, this is consistent with the observed decrease
in continuum flux between the two observations.

\section{Discussion}

\subsection{The Environment of the Active Nucleus in \mrk509}
The velocities for the absorbing gas in \mrk509 span a range from negative to
positive that is atypical for AGN.
When UV absorption lines are detected in low-redshift AGN, they are generally
blue shifted \citep{Crenshaw03, Dunn07}.
This is also true of higher redshift quasars \citep{Weymann79, Ganguly01, Misawa07},
although \citet{Weymann79} did suggest that systems with
red-shifted absorption lines could be associated with neighboring galaxies in
clusters. However, given the significant blue shifts of the UV emission
lines often used to define the quasar redshift, such redshifted systems
could well be blueshifted relative to the systemic velocities of the host
galaxies. In fact, \citet{Ganguly01} finds that the distribution of
associated narrow-line absorbers in QSOs peaks around the UV emission-line
velocity. Nevertheless, one might expect to find gas with random
motions, both blue- and red-shifted, relative to the host galaxy
intercepting our line of sight either
within the host galaxy itself or associated with its neighbors.
Indeed, one component of the absorption in the Seyfert 1 galaxy Mrk 279 has a
positive velocity of $+90~\kms$, and \citet{Scott04a} have suggested that it
is interstellar material in the host galaxy, or gas from a recent encounter
with a close companion, perhaps similar to a high-velocity cloud (HVC) in
our own Milky Way.

\begin{figure}
\centering
\includegraphics[width=8.5cm, scale=0.4]{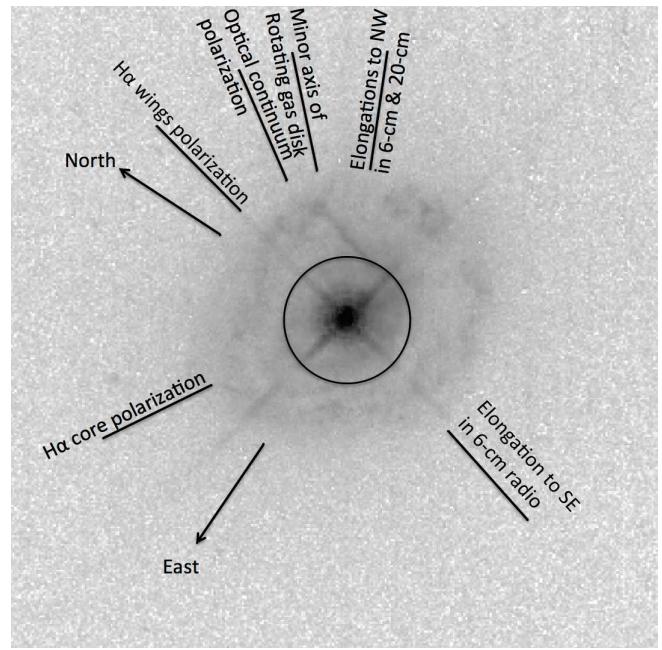}
\caption{
HST WFPC2 image of \mrk509 through filter F547M. The directions North and
East are marked with arrows of length $2.5\arcsec$. The black circle
shows the placement of the $2.5\arcsec$ diameter COS aperture used for
our observations. Lines indicating the position angles of various features
associated with the structure of \mrk509 are indicated. The small knot at
the end of the linear structure visible in the F547M image directly above the
nucleus lies at a
position angle (PA, measured East of North) of $-42\deg$ from the nucleus.
The minor axis (PA$=135\deg$) of the low-ionization gas disk is
from \citet{Phillips83}. The PA of the
polarized light in the core of the H$\alpha$ emission line ($60\deg$), the
broad wings of H$\alpha$ ($168\deg$) and the optical continuum ($144\deg$) are
from \citet{Young99}. The radio structures (PA$= -65\deg$ and $165\deg$)
are from \citet{Singh92}. Note that extended radio emission to the NW rotates
from PA$=-40\deg$ to PA$=-65\deg$ with increasing distance from the nucleus.
}
\label{fig_wfpc2}
\end{figure}

Since our absorption-line observations only
probe gas that is along the line of sight, taking a broader view of the
environment of the active nucleus in \mrk509 should be helpful.
The optical imaging and two-dimensional optical spectroscopy of
\citet{Phillips83} reveal a compact host, perhaps of type S0.
The spectroscopy shows kinematically distinct low-ionization and high-ionization
emission-line components.
The low-ionization gas covers
the face of the galaxy and rotates as a disk about an axis at a
position angle (PA, measured East of North) of $-45\deg$.
The high-ionization gas also extends to more than 10\arcsec\ from the nucleus,
and it is mostly blueshifted, with velocities as high as $-200~\kmps$,
although there are also regions SE of the nucleus with positive velocities.
Radio images of \mrk509 show elongation and extensions to both the NE and
the SW \citep{Singh92}.

To illustrate these features on the spatial scale of our HST observations, we
have retrieved the WFPC2 F547M image of \mrk509 obtained by \citet{Bentz09} from
the MAST archive. We used MultiDrizzle \citep{Fruchter09} to combine the
longest PC exposures (with exposure times of 60 s, 160 s, and 300 s) and
eliminate cosmic rays.
Figure \ref{fig_wfpc2} shows the resultant image (covering the inner 4 kpc of
\mrk509).
Note that this image does not capture the full extent of \mrk509.
The 2.5$\arcsec$-diameter COS aperture covers only the central 870 pc of
\mrk509, while the most prominent features in the HST image, possibly an
irregular starburst ring, have an outer radius of $< 2$ kpc.
If this ring-like feature is due to recent star formation, this may be the
source of ionization for the low-ionization rotating disk in the spectroscopy
of \citet{Phillips83}.
There is also an unusual, linear feature to the NE, which is not aligned with
the nucleus, although it lies in the general direction of the elongations to
the NW in the radio maps.
Deep ground-based CCD images of \mrk509 show an amorphous optical morphology
\citep{MacKenty90} with a broad extension to the SE in the same direction
as the elongation in the 6-cm radio image of \citet{Singh92}.
Although the light is highly concentrated, \citet{MacKenty90} traces an
exponential disk out to a radius of more than 30 kpc.

In addition to this macroscopic view of the nuclear region, optical
spectropolarimetric observations have probed structures that are unresolved
in the optical or radio images.
Optical spectropolarimetry of \mrk509 suggests that there are multiple
scattering regions, each consistent with various aspects of the unified
model of AGN \citep{Young00}.
The optical continuum is polarized at PA=139--154$\deg$, along the same
direction as the inner portions of the 6-cm and 20-cm radio images
\citep{Singh92}. The polarization is also variable \citep{Young99}, and
\citet{Young00} identified this as radiation scattered from the accretion disk.
The \Halpha\ emission line exhibits more complexity in its polarization.
The core is polarized at PA=60$\deg$, nearly perpendicular to the radio jet,
as observed in Type II AGN, where the scattering surface is the mirror of
hot electrons in the cone along the radio axis, perpendicular to the plane
of the obscuring torus.
The wings of \Halpha\ are polarized at PA=168$\deg$, similar to, but not quite
aligned with the inner radio jet.
Figure \ref{fig_wfpc2} shows these orientations relative to other
structures in \mrk509.

As we noted in \S3.4, the COS and FUSE spectra show significantly more
residual light at the bottoms of the intrinsic absorption-line troughs than
does the STIS spectrum. Since COS and FUSE used larger apertures, we
argued that much of this light was scattered light, primarily from a region
less than 0.44445$\arcsec$ in diameter surrounding the active nucleus.
(This scattered light is from a much smaller region than the broad Balmer
emission that \citet{Mediavilla98} detected using two-dimensional
spectroscopy over a region of $\sim8 \arcsec$ surrounding the nucleus,
which they interpret as scattered light from the AGN.)
A distance of 0.225$\arcsec$ corresponds to a radial distance of $< 150$ pc,
which would place it in the typical region associated with reflective
hot-electron mirrors in Type II AGN,
as we noted in \S3.4.
This is consistent with the polarization component seen in the
core of the \Halpha\ emission line that is polarized perpendicularly to the
radio axis. The amount of scattered light we see, approximately 5\% of
the total flux, is also typical of the amount of scattered light seen in
Type II AGN like NGC~1068 \citep{MGM91, Antonucci93}.

Thus our line of sight to the nucleus in \mrk509 may pass through a
complicated mix of components. Based on the kinematics and physical
characteristics of the absorption components revealed by our spectrum,
we will try to link the absorption components to the various features
in \mrk509.

\subsection{Characterizing the Absorbers}

In this paper we take a kinematic approach to identifying and characterizing
the absorbers in \mrk509 (as opposed to the ionization-state approach used
by \citealt{Detmers11} and \citealt{Ebrero11a}).
As in \citet{Ebrero11a}, our discussion is a preliminary assessment of the
properties of the absorbers.
A more comprehensive analysis of the X-ray and UV properties of the
absorbing gas is deferred to a future paper (Ebrero, in preps.).

On the largest velocity scales, the intrinsic absorption in \mrk509 can be
characterized by two main troughs.
The most blue-shifted trough, extending from $-200~\kms$ to $-450~\kms$,
corresponds to the velocities of the most prominent blue-shifted absorption
component in the RGS spectrum ($-319~\kms$; \citealt{Detmers11}) and the
{\it Chandra} spectrum ($-196~\kms$; \citealt{Ebrero11a}).
(This velocity offset between the RGS and LETGS results is not significant,
and it lies within the range of systematic errors in the relative RGS and LETGS
wavelength scales as discussed
by \citealt{Kaastra11b} and \citealt{Ebrero11a}.)
Given the variability we have seen in portions of this trough and its blueshift,
we identify this trough as an outflow from the AGN.

The other main absorption trough in \mrk509 extends from  $-100~\kms$ to
$+250~\kms$. This also matches the kinematics of the other main absorption
component detected in the X-ray spectra, with velocities measured in the
RGS spectrum of $-13~\kms$ \citep{Detmers11} and of $+73~\kms$ in the
{\it Chandra} spectrum \citep{Ebrero11a}.
The ionization level in the X-ray for this component is lower, and this gas
is not in pressure equilibrium with the gas in the blue-shifted absorption
trough \citep{Ebrero11a}. As we discuss below, the components in this trough
may arise in several distinctly different regions of \mrk509.
The higher spectral resolution of our UV spectrum permits us to dissect these
absorption features in greater detail, so we now discuss the properties of
individual components that we have identified in the UV absorption spectrum.

\subsubsection{Components 1a, 1, and 1b}
Component 1a is detected only in the far wing of the \lya\ absorption trough.
There is a distinct depression here that is not seen in the profiles of any
of the other absorption lines, including \ovi.
This suggests that the gas here may be very highly ionized, perhaps representing
the only indication in the UV of a more extensive outflow at higher velocities
that is seen in some of the most highly ionized X-ray absorption features
\citep{Detmers11, Ebrero11a}.

Components 1 and 1b show definite evidence of variability from comparing
the STIS and COS observations.
Both vary in \lya, and
component 1 varies in \nv\ as we discussed in \S3.4.
Although more closely spaced in time,
the FUSE observations do not permit us to set any interesting
limits on the density or location of the gas in these components.
These components may have varied between the two FUSE observations, but not at
a level of significance that permits us to infer densities or distances.
The variations in \nv\ and \lya\ are consistent with ionization changes
in response to a decrease in the continuum flux.
Recombination times for hydrogen are too slow to permit us to set any
interesting limits on the density.
However, \nv\ recombines nearly two orders of magnitude faster, and it provides
greater leverage in sensing the density of the absorbing medium.
Following \citet{KK95} and \citet{Nicastro99},
recombination and ionization time scales
depend not only on the density, but also on the relative populations of the
ionization states involved:
$\rm t_{rec} = (n_i / n_{i+1}) / (n_e \alpha_{rec})$.
Using the photoionization model for the \mrk509 absorbers from
\citet{Kraemer03}, component 1 has an ionization parameter $\rm log~\xi = 0.67$
(where $\xi = \rm L_{ion} / (n r^2)$).
At this level, the ionization fraction for \nv\ is $\sim 0.4$, and the
relative equality of populations among the neighboring ionization states of
nitrogen makes ionization and recombination timescales similar
\citep{Nicastro99}.
For $\alpha_{rec} = 8.96  \times 10^{-12}~cm^3~s^{-1}$ \citep{Nahar06}
at a temperature of 20,000 K, and a time between the STIS and COS observations
of $2.733 \times 10^{8}$ s, we get a lower limit on the density of
$\rm n_e > 160~\cc$. Since we are using the
photoionization models of \citet{Kraemer03}, we use an ionizing luminosity
from their SED in \citet{Yaqoob03},
$\rm L_{ion} = 7.0 \times 10^{44}~\ergsec$.
Together with the ionization parameter $\rm log \xi = 0.67$,
this gives an upper limit on the distance of $\rm r < 250~pc$.

Component 1 is well defined in both \nv\ and \civ, but in \lya\ and \ovi\ it is
merely part of a smoothly varying trough that runs from the deepest point in
components 2 and 2b. As modeled by \citet{Kraemer03}
(who only used the STIS data),
this trough has only
moderate ionization, probably even lower than what they model since they
grossly overpredict the amount of \ovi\ that should be present at this
velocity.
(Their trough at $-420$ \kmps\ is three times deeper than the observed
\ovi\ $\lambda1032$ absorption in the FUSE spectrum at this velocity.)
In fact, as we discussed in \S3.4, given the relative character of the
variability in \lya\ and \nv, the ionization state of this feature is
below the peak in ionization for \nv.
Component 1b appears to have higher ionization since it is strong in \ovi\ but
very indistinct in \nv. Although moderately strong in \lya, it is also weaker
than component 1.
If either of these components are associated with the X-ray portion of the
outflow (which they do not match in detail kinematically), they are likely to be
higher-density, lower-ionization clumps embedded in a more tenuous, highly
ionized outflow.

\subsubsection{Components 2 and 2b}

Components 2 and 2b represent the deepest portions of the blue half of the
absorption trough in the \mrk509 spectrum. Kinematically, these components
are the closest match to one of the major components detected in the
X-ray spectrum of \mrk509.
In the {\it XMM-Newton}/RGS spectrum, component 2 of \citet{Detmers11} is almost an
exact match in velocity to UV component 2 ($-319~\kms$ vs. $-321~\kms$,
respectively).
The ionization state of the gas detected in the UV absorption lines, however, is
not particularly high, and its total column density is less than that seen in
the X-ray. Again, this may be an example of a higher-density, lower-ionization
clump embedded in the more highly ionized outflow seen in the X-ray.

\subsubsection{Component 3}

Component 3 is the closest match in velocity to Component 2 of the {\it Chandra}
LETGS spectrum \citep{Ebrero11a}, but in the UV it too has rather low ionization,
similar in character to UV component 1. As with the previous components, this
gas detected in the UV is likely to be a higher-density clump embedded in
more highly ionized outflowing gas.

\subsubsection{Component 4a}

Component 4a, first identified by \citet{Kraemer03}, shows up most clearly in
\nv\ and \civ. 
In velocity it is close to component 1 identified in the RGS spectrum
by \citet{Detmers11}. Its modest outflow velocity could suggest an association
with the base of the outflowing wind in \mrk509.
However, like all the components above, it is rather low in ionization, and
its column density is lower than that seen in the X-ray absorber.
It is probably not directly associated with the X-ray absorbing gas.

\subsubsection{Component 4}

Perhaps the easiest component to characterize is \#4.
This absorber has the lowest ionization state of any of the components.
It is the only one in which \siiv\ is detected, and it has the strongest
\ciii\ absorption in the original FUSE spectrum of \citet{Kriss00}.
Its velocity of $\sim -22~\rm \kms$ is the closest to our adopted systemic
redshift for \mrk509.
Given the high column densities, relatively low ionization state and velocity
coincidence, it is likely that this component is the interstellar medium of
the \mrk509 host galaxy. 
\citet{Phillips83} find a systemic velocity for the rotating
low-ionization gas disk in their observations that is in good agreement with the
velocity of the {[O\,{\sc iii}]} line-center velocity which they adopted as the systemic
velocity of the host galaxy. To within their errors of $\pm 30~\rm \kms$,
all of these velocities are consistent with the velocity we measure
for component 4.
\citet{Kraemer03} find that the broad range of UV ionization states present at
this velocity requires more than one component in a photoionization model
solution.
Component 4 has the closest match in velocity to component 1 identified in
the RGS spectrum by \citet{Detmers11}.
Within the errors of the {\it Chandra} spectrum, Component 4 may also be associated
with component 1 from the LETGS spectrum \citep{Ebrero11a}.
Given its low ionization parameter, \citet{Ebrero11a} suggest that this
X-ray absorption may be associated with the ISM or halo of the host galaxy of
\mrk509.

\subsubsection{Component 5}

Component 5 is similar in character to Component 4, but it has lower overall
column density and slightly higher ionization.
Its slightly positive velocity is a good match to component 1 in the LETGS
spectrum. Highly ionized gas at this velocity is also seen in the
three-dimensional spectroscopy of \citet{Phillips83}, and it is within the
velocity profile of the echelle spectrum of {[O\,{\sc iii}]} in the nuclear
region of \mrk509. It appears to have varied in strength in \hi\ between the
two FUSE observations. As it is difficult to disentangle this component's
variations from those of components 6a, 6, and 6b, we discuss the implications
of this variation in the next subsection.

\subsubsection{Components 6a, 6, and 6b}

Although component 6 has a positive velocity, this does not preclude it from
an association with the active nucleus in \mrk509.
In fact, if one examines the {[O\,{\sc iii}]} echelle spectrum of the nuclear
region in \mrk509 in Fig. 3 of \citet{Phillips83}, one can see that the actual
peak of the line is at a positive velocity of $+120~\rm \kms$, matching our
velocity for absorption in component 6.
The kinematics of X-ray-heated winds modeled by \citet{Balsara93} shows that
portions of the flow that are evaporated off the inner edge of the
obscuring torus can be captured by the central black hole.
This results in streamlines with positive, inflow velocities.

The variability in \hi\ absorption we see over the 10 months
between the two FUSE observations suggests a close relationship to the
active nucleus for components 5 and 6,
and, as for Component 1, it allows us to set an upper limit on the
distance of this gas from the central engine.
From 1999 November to 2000 September, over an interval of 303 days,
the continuum
flux from \mrk509 dropped by 30\%, and the hydrogen column density in
components 5 and 6 increased by 14\%, as we showed in \S3.4.
This gives us an upper limit on the recombination time for the gas, and hence
a lower limit on the density.
Again using the photoionization model from
\citet{Kraemer03}, component 6 has an ionization parameter $\rm log~\xi = 0.71$,
and an ionization fraction for hydrogen of $2.3 \times 10^{-5}$.
For a time interval between the two FUSE observations of $2.62 \times 10^{7}$ s,
and $\rm \alpha_{rec} = 1.43 \times 10^{-13}~cm^3~s^{-1}$ \citep{Osterbrock}
at a temperature of 20,000 K, we get a limit on the density of
$\rm n_e > 6.0~\cc$. 
The ionizing luminosity of $\rm L_{ion} = 7.0 \times 10^{44}~\ergsec$
together with the ionization parameter $\rm log \xi = 0.71$,
gives an upper limit on the distance of $\rm r < 1.5~kpc$.
So, component 6 is definitely within the confines of the host galaxy, and
likely in the near vicinity of the nucleus.

\subsubsection{Components 7a and 7}

Component 7 at +219 $\kms$, from its velocity, is not outflowing, but
infalling to \mrk509.
Component 7a is strongly blended with \#7 in \ovi\ and \lya;
it only shows up as a kinematically independent feature in \nv.
Neither component shows any significant variability either between
the two FUSE observations in 1999 and 2000, or between the longer interval
between STIS and COS from 2001 to 2009.
Given that it is optically thin in \nv, \civ, and \lya, this lack of
variability is not being masked by saturation effects that limit our ability
to see variations in other components.
Based on just the qualitative appearance of the UV spectra, this is the highest
ionization parameter trough in the whole spectrum---\ovi\ is very strong,
\nv\ is present, and \civ\ is not detectable.
The stacked RGS spectrum at these velocities \citep{Detmers11}
does not reveal absorption in \ovi\ K$\alpha$,
although the upper limits are consistent with the FUSE spectrum.
So, even though the ionization level exhibited in the UV is high, it is still
not as high as the blue-shifted trough in the X-ray spectrum.
\citet{Kraemer03} and \citet{Kriss00} found their highest ionization
parameters for this component as well.

We suggest that components 7 and 7a might be similar to HVC complexes
seen in our own galaxy, except that the ionization of these clouds in \mrk509
would be completely dominated by the radiation from the active nucleus rather
than the extragalactic background radiation and starlight from the galaxy.
These clouds may be infalling extragalactic material, or they may be tidal
material stripped from an encounter with a companion in the past. Although
\mrk509 does not have an obvious interacting companion, its disturbed outer
isophotes \citep{MacKenty90} suggest there may have been an encounter in the
more distant past.
\citet{Thom08} find that HVC Complex C in the Milky Way
has an average density $\rm log~n \sim -2.5$, dimensions $3 \times 15$ kpc,
a distance of $10 \pm 2.5$ kpc, and a total mass of
$8.2^{+4.6}_{-2.6} \times 10^6~\Msun$.
If components 7 and 7a in \mrk509 have roughly similar densities of
$n = 0.01~\cc$, then we can use the photoionization results of
\citet{Kraemer03} to estimate their size and distance
from the nucleus of \mrk509.
\citet{Kraemer03} give a total column density of
$N_H = 4 \times 10^{19}~\pcm2$ for component 7, which implies a size
(as a cube) of $\rm 1.3 (0.01~\cc/n)$ kpc.
For their quoted ionization parameter of $\rm log~\xi = 1.33$,
and an ionizing luminosity of $\rm L_{ion} = 7.0 \times 10^{44}~\ergsec$,
the distance from the center of \mrk509 is $19~(0.01~\cc/n)^{0.5}$ kpc.
This is just about right for a structure comparable to a Milky Way HVC.
For a density this low and ionization this high, the recombination time for
\hi\ in components 7a and 7 would be $80(0.01~\cc/n)$ years, so the
lack of variability in our observations is consistent with this hypothesis.

\subsubsection{Smoothly distributed gas in the outflow}

While we have characterized the absorption features in \mrk509 in terms of
discrete kinematic components, the smoothly varying profiles of the \lya\ and
\ovi\ absorption in the blue-shifted outflow trough 
(see Fig. \ref{fig_cosvel}) suggest that additional gas may be present
that is smoothly distributed over a range in velocities.
The tail to higher velocities observed as component 1a in \lya\ may be a part
of this distribution.
If this smoothly distributed gas is very highly ionized, one might expect that
the only traces remaining in the UV absorption spectrum would be the most highly
ionized species, like \ovi, or the most abundant ones, like \hi.
This more smoothly distributed, highly ionized gas could be the counterpart of
the features detected in the RGS and LETGS X-ray spectra.

The discrete UV absorption components we have identified
show up most clearly in \nv\ and \civ.
As noted in \S3.3, these often do not precisely align with
similar features in \ovi\ and \lya. For example, in the region surrounding
component 1b, there is little \civ\ and \nv\ absorption, but moderate depth
in \ovi\ and the hydrogen lines, although it is a local minimum in \hi.
The lower ionization, more discrete components we observe in \nv\ and \civ\ 
could be dense, lower-ionization clumps embedded in this flow,
or clouds in the interstellar
medium of \mrk509 that are being entrained in the flow.
As shown by \citet{Ebrero11a}, these lower-ionization components are not in
pressure equilibrium with the other X-ray absorbing components.
Such discrete clouds with similar kinematics are spatially resolved and
observed in the nearest AGN, e.g., NGC 4151 \citep{Das05} and
NGC 1068 \citep{Das06, Das07}.
This outflowing gas is also visible in the three dimensional spectroscopy
of \citet{Phillips83}.

\subsection{Thermal Winds vs. Disk Winds}

Disk winds are a popular explanation for outflows from AGN
\citep{Konigl94, Murray95, Elvis00, Proga00, Richards10, Fukumura10}.
For the high velocities observed in broad-absorption-line (BAL) QSOs, an
origin for the wind from the accretion disk deep in the potential well
of the central black hole is a natural inference.
For the low velocities we observe in \mrk509, it is not so clear.
\citet{Kraemer03} summarizes the problems faced in trying to explain the
outflow in \mrk509 with a disk wind.
Briefly, \mrk509 is likely to be observed nearly face on to the accretion disk
based on 
the polarized flux in the broad wings of \Halpha\ that is
polarized parallel to the radio jet, suggesting an origin in scattering from
the face of the accretion disk \citep{Young99, Young00}.
Since disk winds roughly lie in the plane of the disk, the flow would be
predominantly perpendicular to our line of sight.
This can explain the low velocities since most of the motion would be
transverse to our line of sight, but it makes it difficult to explain the
near-unity covering factors of the absorbers if the UV continuum is
produced interior to the launch point of the wind.

The modest velocities of hundreds of $\kms$ we observe here are more
characteristic of thermal winds produced in the AGN environment.
Irradiation of the obscuring torus in an AGN by the nuclear flux
produces material that fills the cone above the torus and flows outward at
velocities of hundreds of $\kms$, typical of the sound speed in the gas
\citep{Balsara93, KK95, KK01}.
The presence of some redshifted components is also consistent with portions
of an X-ray-heated wind that are captured by the central black hole
\citep{Balsara93}.
These winds are also expected to have multiple phases of gas as the
surface of the torus is ablated by the radiation from the central engine
\citep{KK95, KK01}.
The most highly ionized portions of this flow can produce the X-ray
absorption.
The dense part of the wind near the base of the ionization cone may be the
slightly extended scattering region producing the extra flux we see in the
bottoms of the absorption troughs and also produce the polarization seen in the
core of the \Halpha\ emission line.
As we noted in the prior section, the lower-ionization UV absorption
components could be denser clumps embedded in this flow.
The extended high-ionization {[O\,{\sc iii}]} emission outflowing from the
nuclear region and studied by \citet{Phillips83} could be analogous to the
discrete clouds imaged with HST in nearby AGN that show similar
kinematics.

\section{Conclusions}

We have presented HST/COS observations of the Seyfert 1 galaxy \mrk509
obtained as part of an extensive multiwavelength campaign \citep{Kaastra11a}
on this galaxy in late 2009.
Our UV spectra were obtained simultaneously with
{\it Chandra} LETGS spectra \citep{Ebrero11a}.
The campaign also included immediately prior observations with
{\it XMM-Newton}, Swift, and INTEGRAL.
Our spectra cover the 1155--1760~\AA\ wavelength range at a resolution of
$\sim15~\kms$, and they are the highest signal-to-noise observations to date of
the intrinsic absorption components of \mrk509.
This enables us to trace additional complexity in the absorption troughs
compared to prior STIS observations \citep{Kraemer03}.
As part of our analysis, we have also examined archival FUSE observations,
which provide additional information on variability characteristics of the
absorbers.

In our COS spectra we identify a total of 14 kinematic components.
Six of these are blends that represent subcomponents
of previously known features.
The most blueshifted portions of the \lya\ absorption trough
(components 1 and 1b) show a
significant decrease in column density compared to prior STIS observations.
At the same time, component 1 in \nv\ increased in strength compared to the
STIS observations.
Using recombination timescale arguments, the variation in the \nv\ column
density allows us to set a lower limit on the gas density of $>160~\cc$, and
an upper limit on the distance of $< 250$ pc
for this most blue-shifted component.
We also detect variability between the two FUSE observations from
1999 November \citep{Kriss00} to 2000 September in component 6 at a positive
velocity of $+120~\kms$, which corresponds to the peak of the {[O\,{\sc iii}]}
emission-line profile from the echelle spectrum of \citet{Phillips83}.
Based on the recombination time, we set a limit on the density of this
component of $>6~\cc$, and set an upper limit on its distance from the
nucleus at $<1.5~\rm kpc$.

The COS spectra also show greater residual light at the bottoms of the
absorption troughs relative to the STIS spectrum.
The fraction of residual light seen in COS is comparable to that seen in
the FUSE spectra.
We attribute this residual light to scattering from an extended region near
the nucleus, which is consistent with the larger apertures used for the
COS and FUSE observations relative to the STIS spectrum.
Analysis of the cross-dispersion profile in the COS spectrum limits the
spatial extent of this scattering region to $< 0.5 \arcsec$, or
$< 170$ pc.
Scattering from such a region would be consistent with electron scattering from
the base of an ionization cone in \mrk509, and it would explain the
polarization in the core of \Halpha\ \citep{Young99} which is observed to be
perpendicular to the radio jet \citep{Singh92}.

The velocities of the two main absorption troughs roughly correspond to the
kinematics of the detected X-ray absorption.
The UV absorption in \mrk509 arises from a variety of sources.
The lowest-velocity absorption trough covers a velocity range of $-100~\kms$ to
$+250~\kms$.
The deepest portion of this trough in the UV is at the systemic velocity of
\mrk509, and it has the lowest ionization.
This is also true for the X-ray absorption. 
We attribute this portion of the absorption
to the interstellar medium or galactic halo of \mrk509.
The most redshifted portion of the trough has characteristics comparable to
the high-velocity cloud Complex C in our own Milky Way \citep{Thom08}.
At a density of $0.01~\cc$ (which we can only assume, not measure),
it would have a size of roughly 1.3 kpc at a distance of 19 kpc from the center
of \mrk509.

The most blue-shifted absorption trough in \mrk509 is
an outflow from the active nucleus.
The velocities in this trough of $-200~\kms$ to $-450~\kms$ correspond to the
most highly ionized portion of the X-ray absorbing gas
\citep{Detmers11, Ebrero11a},
and they overlap with the extended emission from high-ionization gas
that covers the face of \mrk509 in the three-dimensional spectroscopy of
\citet{Phillips83}.
The outflow velocities, embedded clumps of lower ionization gas,
the presence of an extended scattering region, and the extended outflowing
emission seen in {[O\,{\sc iii}]} \citep{Phillips83} are all
consistent with an origin in a multiphase thermal wind produced by the
irradiation of an obscuring torus by the active nucleus.

\begin{acknowledgements}
This work was supported by NASA through grants for HST program number 12022
from the Space Telescope Science Institute, which is operated
by the Association of Universities for Research in Astronomy, Incorporated,
under NASA contract NAS5-26555, 
{\it XMM-Newton} grant NNX09AR01G from Goddard Space Flight Center.
SRON is supported financially by NWO,
the Netherlands Organization for Scientific Research.
KCS acknowledges the support of Comit\'e Mixto ESO - Gobierno de Chile.
SB acknowledges financial support from contract ASI-INAF n. I/088/06/0.
EB was supported by a grant from the Israel Science Foundation.
PoP acknowledges financial support from the GDR PCHE in France and
from the CNES French national space agency.
GP acknowledges support via an EU Marie Curie Intra-European Fellowship under
contract no. FP7-PEOPLE-2009-IEF-254279.
\end{acknowledgements}

\bibliographystyle{aa}
\bibliography{17123}


\setcounter{table}{2}
\begin{table*}
	\caption[]{Emission Features in the COS Spectrum of \mrk509}
	\label{COS_elines}
\begin{tabular}{l c c c c}
\hline\hline       
Feature & $\rm \lambda_0$ & Flux & $\rm v_{sys}$ & FWHM \\
  & ($\rm \AA$)  & ($\rm 10^{-14}~erg~cm^{-2}~s^{-1}~\AA^{-1}$) & ($\rm km~s^{-1}$) & ($\rm km~s^{-1}$) \\
\hline
\lya  & 1215.670 & $   9.1\pm  2.5$ & $   166 \pm  31$ & $  300 \pm   23$\\
\lya  & 1215.670 & $  73.9\pm  2.5$ & $  -151 \pm  15$ & $ 1339 \pm   33$\\
\lya  & 1215.670 & $ 410.0\pm  1.8$ & $  -235 \pm   8$ & $ 3039 \pm   14$\\
\lya  & 1215.670 & $ 838.0\pm  1.8$ & $  -156 \pm   7$ & $ 9677 \pm   26$\\
\nv   & 1238.821 & $  22.1\pm  0.6$ & $   205 \pm  22$ & $ 2164 \pm   45$\\
\nv   & 1242.804 & $  11.1\pm  0.2$ & $   205 \pm  22$ & $ 2164 \pm   45$\\
\siii  & 1260.420 & $  65.3\pm  1.0$ & $  -707 \pm  36$ & $ 5652 \pm   66$\\
O~{\sc i}+Si~{\sc ii} & 1303.57 & $  31.5\pm  0.6$ & $   36 \pm  33$ & $ 3307 \pm   90$\\
\cii   & 1335.205 & $   8.0\pm  1.5$ & $     8 \pm  67$ & $ 2176 \pm  424$\\
\siiv  & 1393.755 & $  85.0\pm  0.2$ & $   492 \pm   9$ & $ 5084 \pm   36$\\
\siiv  & 1402.770 & $  42.5\pm  0.1$ & $   491 \pm   9$ & $ 5084 \pm   36$\\
\civ   & 1548.195 & $   4.2\pm  1.4$ & $  -204 \pm  39$ & $  300 \pm   39$\\
\civ   & 1550.770 & $   2.1\pm  0.3$ & $  -204 \pm  39$ & $  300 \pm   39$\\
\civ   & 1548.195 & $  78.5\pm  0.5$ & $  -206 \pm   7$ & $ 1974 \pm   15$\\
\civ   & 1550.770 & $  39.3\pm  0.5$ & $  -206 \pm   7$ & $ 1974 \pm   15$\\
\civ   & 1548.195 & $ 199.0\pm  1.9$ & $  -206 \pm   7$ & $ 4233 \pm   15$\\
\civ   & 1550.770 & $  99.3\pm  1.9$ & $  -206 \pm   7$ & $ 4233 \pm   15$\\
\civ   & 1549.050 & $ 461.0\pm  1.1$ & $   158 \pm  10$ & $10299 \pm   56$\\
N~{\sc iv}] & 1486.496 & $ 7.0\pm  0.1$ & $ -33 \pm 61$ & $ 2181 \pm  108$\\
\siii  & 1526.710 & $  15.7\pm  1.3$ & $     2 \pm  27$ & $ 2181 \pm  108$\\
\heii & 1640.480 & $   8.5\pm  0.3$ & $  -113 \pm  39$ & $ 1153 \pm   83$\\
\heii & 1640.480 & $  55.4\pm  0.5$ & $  -199 \pm  25$ & $ 4065 \pm   61$\\
O~{\sc iii}] & 1663.445 & $ 66.0\pm 1.6$ & $ 83 \pm 19$ & $ 3900 \pm   69$\\
\hline                  
\end{tabular}
\end{table*}

\begin{table*}
	\caption[]{Emission Features in the STIS Spectrum of \mrk509}
	\label{STIS_elines}
\begin{tabular}{l c c c c}
\hline\hline       
Feature & $\rm \lambda_0$ & Flux & $\rm v_{sys}$ & FWHM \\
  & ($\rm \AA$)  & ($\rm 10^{-14}~erg~cm^{-2}~s^{-1}~\AA^{-1}$) & ($\rm km~s^{-1}$) & ($\rm km~s^{-1}$) \\
\hline
\lya      & 1215.6700 & $   6.8\pm  5.5$ & $   166 \pm 575$ & $  300 \pm    3$\\
\lya      & 1215.6700 & $ 102.0\pm  0.9$ & $    85 \pm  15$ & $ 1712 \pm   66$\\
\lya      & 1215.6700 & $ 350.0\pm  2.9$ & $  -290 \pm   4$ & $ 3483 \pm   25$\\
\lya      & 1215.6700 & $ 531.0\pm  4.4$ & $   -11 \pm  19$ & $10752 \pm   56$\\
\nv       & 1238.8210 & $  17.4\pm  0.3$ & $   371 \pm  44$ & $ 2089 \pm   39$\\
\nv       & 1242.8040 & $   8.7\pm  0.2$ & $   371 \pm  44$ & $ 2089 \pm   39$\\
\siii      & 1260.4200 & $  36.1\pm  1.0$ & $  -921 \pm 130$ & $ 5928 \pm  151$\\
O~{\sc i}+Si~{\sc ii}   & 1303.5700 & $  21.8\pm  2.5$ & $   189 \pm  62$ & $ 3115 \pm   94$\\
\cii      & 1335.2050 & $   4.7\pm  2.4$ & $   106 \pm 284$ & $ 1555 \pm  459$\\
\siiv     & 1393.7550 & $  54.9\pm  3.0$ & $   381 \pm 194$ & $ 4374 \pm   97$\\
\siiv     & 1402.7700 & $  27.4\pm  1.5$ & $   382 \pm 194$ & $ 4374 \pm   97$\\
\civ      & 1548.1950 & $   6.7\pm  1.8$ & $  -204 \pm 151$ & $  300 \pm   95$\\
\civ      & 1550.7700 & $   3.4\pm  0.9$ & $  -204 \pm 151$ & $  300 \pm   95$\\
\civ      & 1548.1950 & $  65.9\pm  3.6$ & $  -103 \pm   7$ & $ 2108 \pm   30$\\
\civ      & 1550.7700 & $  33.0\pm  1.8$ & $  -103 \pm   7$ & $ 2108 \pm   30$\\
\civ      & 1548.1950 & $ 148.0\pm  2.0$ & $  -103 \pm   7$ & $ 4102 \pm   93$\\
\civ      & 1550.7700 & $  74.1\pm  1.0$ & $  -103 \pm   7$ & $ 4102 \pm   93$\\
\civ      & 1549.0530 & $ 389.0\pm  1.5$ & $  -605 \pm  10$ & $ 9392 \pm  156$\\
N~{\sc iv}] & 1486.4960 & $  12.3\pm  0.6$ & $   262 \pm  71$ & $ 3795 \pm   20$\\
\siii      & 1526.7070 & $  15.8\pm  1.4$ & $  -175 \pm  86$ & $ 3795 \pm   20$\\
\heii     & 1640.4800 & $   5.4\pm  0.8$ & $    38 \pm  48$ & $  888 \pm  135$\\
\heii     & 1640.4800 & $  39.5\pm  3.1$ & $  -636 \pm  67$ & $ 4824 \pm  104$\\
\hline                  
\end{tabular}
\end{table*}

\begin{table*}
	\caption[]{Emission Features in the 1999 FUSE Spectrum of \mrk509}
	\label{1999FUSE_elines}
\begin{tabular}{l c c c c}
\hline\hline       
Feature & $\rm \lambda_0$ & Flux & $\rm v_{sys}$ & FWHM \\
  & ($\rm \AA$)  & ($\rm 10^{-14}~erg~cm^{-2}~s^{-1}~\AA^{-1}$) & ($\rm km~s^{-1}$) & ($\rm km~s^{-1}$) \\
\hline
\ciii     &  977.0200 & $  60.0\pm  4.7$ & $    -8 \pm  25$ & $ 4000 \pm  172$\\
\niii     &  989.7990 & $  56.1\pm  1.8$ & $  -546 \pm  51$ & $ 8917 \pm  217$\\
\lyb      & 1025.7220 & $   0.6\pm  5.1$ & $  -812 \pm 197$ & $ 8917 \pm  217$\\
\ovi       & 1031.9260 & $ 100.0\pm  3.2$ & $  -812 \pm 197$ & $ 8917 \pm  217$\\
\ovi       & 1037.6170 & $  50.1\pm  1.6$ & $  -812 \pm 197$ & $ 8917 \pm  217$\\
\ovi       & 1031.9260 & $  69.0\pm  3.1$ & $   193 \pm  48$ & $ 4182 \pm   99$\\
\ovi       & 1037.6170 & $  34.5\pm  1.6$ & $   193 \pm  48$ & $ 4182 \pm   99$\\
\ovi       & 1031.9260 & $   3.4\pm  0.6$ & $  -203 \pm  33$ & $  300 \pm   47$\\
\ovi       & 1037.6170 & $   1.7\pm  0.3$ & $  -203 \pm  33$ & $  300 \pm   47$\\
\hline                  
\end{tabular}
\end{table*}

\begin{table*}
	\caption[]{Emission Features in the 2000 FUSE Spectrum of \mrk509}
	\label{2000FUSE_elines}
\begin{tabular}{l c c c c}
\hline\hline       
Feature & $\rm \lambda_0$ & Flux & $\rm v_{sys}$ & FWHM \\
  & ($\rm \AA$)  & ($\rm 10^{-14}~erg~cm^{-2}~s^{-1}~\AA^{-1}$) & ($\rm km~s^{-1}$) & ($\rm km~s^{-1}$) \\
\hline
\ciii     &  977.0200 & $  26.8\pm  1.1$ & $   858 \pm 665$ & $ 9157 \pm 860$\\
\ciii     &  977.0200 & $   2.5\pm  0.1$ & $   137 \pm 105$ & $  729 \pm 277$\\
\niii     &  989.7990 & $  23.0\pm  1.6$ & $  -297 \pm 245$ & $ 7844 \pm 187$\\
\lyb      & 1025.7220 & $   0.1\pm  1.2$ & $  -566 \pm 245$ & $ 7844 \pm 187$\\
\ovi       & 1031.9260 & $ 109.0\pm  5.0$ & $  -565 \pm 245$ & $ 7844 \pm 187$\\
\ovi       & 1037.6170 & $  54.3\pm  2.5$ & $  -564 \pm 245$ & $ 7844 \pm 187$\\
\ovi       & 1031.9260 & $  43.2\pm  4.0$ & $   199 \pm 855$ & $ 2585 \pm 272$\\
\ovi       & 1037.6170 & $  21.6\pm  2.0$ & $   199 \pm 855$ & $ 2585 \pm 272$\\
\ovi       & 1031.9260 & $   4.2\pm  1.0$ & $  -203 \pm 129$ & $  300 \pm 250$\\
\ovi       & 1037.6170 & $   2.1\pm  0.5$ & $  -203 \pm 129$ & $  300 \pm 250$\\
\hline                  
\end{tabular}
\end{table*}

\newpage

\setcounter{table}{8}
\begin{table*}
	\caption[]{Intrinsic Absorption Features in the COS Spectrum of \mrk509\label{COS_alines}}
\begin{tabular}{l c c c c c}
\hline\hline       
{Ion} & {Component} & {$\rm N$} & {$\rm v_{sys}$} & {$\rm b$} & {$\rm f_c$} \\
      & { \# } & {($\rm 10^{12}~cm^{-2}$) } & {($\rm km~s^{-1}$)} & {($\rm km~s^{-1}$)} &  \\
\hline
\hi   & 1a & $   6.3\pm  0.7$ & $  -426 \pm   5$ & $   13 \pm  1$ & $0.802 \pm 0.087$\\
\hi   & 1  & $  49.8\pm  4.9$ & $  -385 \pm   7$ & $   30 \pm  1$ & $0.802 \pm 0.087$\\
\hi   & 1b & $   2.2\pm  2.1$ & $  -361 \pm   8$ & $   12 \pm  2$ & $0.802 \pm 0.087$\\
\hi   & 2  & $ 349.7\pm 87.5$ & $  -316 \pm   6$ & $   25 \pm  1$ & $0.965 \pm 0.084$\\
\hi   & 2b & $ 133.2\pm 66.1$ & $  -283 \pm   5$ & $   26 \pm  3$ & $0.965 \pm 0.084$\\
\hi   & 3  & $  75.6\pm  9.2$ & $  -240 \pm   7$ & $   32 \pm  1$ & $0.843 \pm 0.126$\\
\hi   & 4a & $ 262.0\pm  9.1$ & $   -59 \pm   5$ & $   29 \pm  1$ & $0.953 \pm 0.005$\\
\hi   & 4  & $ 134.9\pm 27.4$ & $    -8 \pm   6$ & $   31 \pm  1$ & $0.984 \pm 0.028$\\
\hi   & 5  & $ 217.3\pm 30.0$ & $    44 \pm   8$ & $   42 \pm  5$ & $0.970 \pm 0.024$\\
\hi   & 6a & $  34.6\pm 10.0$ & $   132 \pm   7$ & $   33 \pm  2$ & $0.993 \pm 0.024$\\
\hi   & 6  & $ 312.5\pm 35.6$ & $   125 \pm   6$ & $   33 \pm  2$ & $0.993 \pm 0.024$\\
\hi   & 6b & $   0.1\pm  1.0$ & $   166 \pm   8$ & $   33 \pm  2$ & $0.993 \pm 0.024$\\
\hi   & 7a & $  53.8\pm 12.8$ & $   185 \pm   5$ & $   16 \pm  3$ & $0.861 \pm 0.051$\\
\hi   & 7  & $ 123.9\pm  2.5$ & $   220 \pm   5$ & $   32 \pm  2$ & $0.861 \pm 0.051$\\
\nv   & 1  & $ 107.0\pm  9.5$ & $  -408 \pm   5$ & $   10 \pm  1$ & $0.712 \pm 0.026$\\
\nv   & 1b & $  17.5\pm  1.6$ & $  -375 \pm   7$ & $   21 \pm  3$ & $0.678 \pm 0.010$\\
\nv   & 2  & $ 149.0\pm 10.7$ & $  -321 \pm   5$ & $   13 \pm  1$ & $0.950 \pm 0.017$\\
\nv   & 2b & $ 130.6\pm 11.3$ & $  -291 \pm   6$ & $   19 \pm  2$ & $0.729 \pm 0.025$\\
\nv   & 3  & $  88.6\pm  8.1$ & $  -244 \pm   5$ & $   16 \pm  1$ & $0.698 \pm 0.045$\\
\nv   & 4a & $  93.6\pm  6.8$ & $   -59 \pm   5$ & $   13 \pm  1$ & $0.945 \pm 0.031$\\
\nv   & 4  & $ 323.6\pm 12.2$ & $   -19 \pm   5$ & $   14 \pm  1$ & $1.000 \pm 0.012$\\
\nv   & 5  & $  38.3\pm 11.1$ & $    37 \pm   5$ & $   13 \pm  1$ & $1.000 \pm 0.250$\\
\nv   & 6a & $  20.6\pm  6.4$ & $    79 \pm   7$ & $   41 \pm 12$ & $1.000 \pm 0.023$\\
\nv   & 6  & $  56.1\pm  4.0$ & $   121 \pm   5$ & $   13 \pm  1$ & $1.000 \pm 0.023$\\
\nv   & 6b & $   5.5\pm  1.8$ & $   147 \pm   8$ & $   11 \pm  3$ & $1.000 \pm 0.023$\\
\nv   & 7  & $  21.0\pm  2.9$ & $   184 \pm   6$ & $   16 \pm  1$ & $0.836 \pm 0.094$\\
\nv   & 7a & $  23.5\pm  2.7$ & $   222 \pm   6$ & $   16 \pm  1$ & $0.836 \pm 0.094$\\
\siiv & 4  & $   2.5\pm  0.2$ & $   -22 \pm   6$ & $   10 \pm  1$ & $1.000 \pm 0.146$\\
\civ  & 1  & $  31.2\pm  1.5$ & $  -406 \pm   6$ & $   14 \pm  1$ & $0.892 \pm 0.026$\\
\civ  & 1b & $  16.3\pm  6.1$ & $  -365 \pm   5$ & $   14 \pm  1$ & $0.935 \pm 0.261$\\
\civ  & 2  & $ 136.3\pm 41.7$ & $  -320 \pm   6$ & $   13 \pm  1$ & $0.979 \pm 0.035$\\
\civ  & 2b & $ 128.9\pm 41.4$ & $  -292 \pm   7$ & $   22 \pm  3$ & $0.896 \pm 0.030$\\
\civ  & 3  & $  44.7\pm  7.2$ & $  -245 \pm   5$ & $   22 \pm  2$ & $0.987 \pm 0.023$\\
\civ  & 4a & $  66.3\pm  2.1$ & $   -60 \pm   5$ & $   12 \pm  1$ & $0.961 \pm 0.031$\\
\civ  & 4  & $ 250.0\pm 11.0$ & $   -18 \pm   5$ & $   16 \pm  1$ & $0.967 \pm 0.007$\\
\civ  & 5  & $  36.2\pm 16.9$ & $    34 \pm   5$ & $   19 \pm  1$ & $0.939 \pm 0.109$\\
\civ  & 6a & $   5.6\pm  2.3$ & $    88 \pm   7$ & $   12 \pm  2$ & $0.986 \pm 0.049$\\
\civ  & 6  & $  12.3\pm  6.4$ & $   122 \pm   5$ & $   10 \pm  1$ & $0.986 \pm 0.049$\\
\civ  & 6b & $  17.6\pm  5.3$ & $   120 \pm   8$ & $   24 \pm  4$ & $0.986 \pm 0.049$\\
\civ  & 7a & $   3.6\pm  1.1$ & $   171 \pm   9$ & $   13 \pm  2$ & $0.892 \pm 0.126$\\
\civ  & 7  & $   6.4\pm  0.8$ & $   232 \pm   6$ & $   13 \pm  2$ & $0.892 \pm 0.126$\\
\hline                  
\end{tabular}
\\
{The scattered light component in this fit is $0.05 \pm 0.01$.}
\end{table*}

\begin{table*}
	\caption[]{Intrinsic Absorption Features in the STIS Spectrum of \mrk509\label{STIS_alines}}
\begin{tabular}{l c c c c c}
\hline\hline       
{Ion} & {Component} & {$\rm N$} & {$\rm v_{sys}$} & {$\rm b$} & {$\rm f_c$} \\
      & { \# } & {($\rm 10^{12}~cm^{-2}$) } & {($\rm km~s^{-1}$)} & {($\rm km~s^{-1}$)} &  \\
\hline
\hi   & 1a & $  11.6\pm  4.2$ & $  -428 \pm   7$ & $   17 \pm  2$ & $0.561 \pm 0.052$\\
\hi   & 1  & $  63.9\pm 14.5$ & $  -380 \pm   6$ & $   32 \pm  5$ & $0.561 \pm 0.052$\\
\hi   & 1b & $  11.9\pm  9.2$ & $  -359 \pm   6$ & $   14 \pm  6$ & $0.905 \pm 0.052$\\
\hi   & 2  & $ 459.2\pm 276.4$ & $  -314 \pm   7$ & $   24 \pm  6$ & $0.922 \pm 0.019$\\
\hi   & 2b & $  86.8\pm 73.1$ & $  -283 \pm  16$ & $   27 \pm 12$ & $0.850 \pm 0.198$\\
\hi   & 3  & $  87.5\pm 41.8$ & $  -237 \pm   8$ & $   34 \pm  4$ & $0.861 \pm 0.101$\\
\hi   & 4a & $ 250.0\pm 98.3$ & $   -66 \pm   6$ & $   30 \pm  3$ & $0.960 \pm 0.010$\\
\hi   & 4  & $ 180.3\pm 38.0$ & $   -16 \pm   9$ & $   18 \pm  6$ & $0.979 \pm 0.026$\\
\hi   & 5  & $ 351.0\pm 75.2$ & $    37 \pm   6$ & $   30 \pm  7$ & $0.958 \pm 0.049$\\
\hi   & 6a & $ 442.3\pm 188.9$ & $   137 \pm   9$ & $   43 \pm  2$ & $0.998 \pm 0.051$\\
\hi   & 6  & $ 401.4\pm 709.1$ & $   137 \pm  26$ & $   43 \pm  2$ & $0.998 \pm 0.051$\\
\hi   & 6b & $   6.8\pm 34.6$ & $   168 \pm 152$ & $   43 \pm  2$ & $0.998 \pm 0.051$\\
\hi   & 7a & $  75.5\pm 16.2$ & $   187 \pm 134$ & $   19 \pm 94$ & $0.632 \pm 0.277$\\
\hi   & 7  & $ 206.7\pm 28.6$ & $   220 \pm   7$ & $   34 \pm  4$ & $0.632 \pm 0.277$\\
\nv   & 1  & $  95.5\pm 34.1$ & $  -408 \pm   6$ & $   10 \pm  1$ & $0.562 \pm 0.094$\\
\nv   & 1b & $  16.6\pm 29.8$ & $  -375 \pm   6$ & $   21 \pm 12$ & $0.810 \pm 0.104$\\
\nv   & 2  & $ 238.9\pm 56.2$ & $  -319 \pm   7$ & $   12 \pm  2$ & $0.860 \pm 0.072$\\
\nv   & 2b & $  91.7\pm 79.6$ & $  -291 \pm   6$ & $   16 \pm  5$ & $0.809 \pm 0.263$\\
\nv   & 3  & $  94.0\pm 34.3$ & $  -246 \pm   6$ & $   18 \pm  3$ & $0.759 \pm 0.205$\\
\nv   & 4a & $  89.8\pm 21.6$ & $   -57 \pm   6$ & $   14 \pm  1$ & $0.919 \pm 0.133$\\
\nv   & 4  & $ 263.1\pm 23.3$ & $   -17 \pm   6$ & $   17 \pm  1$ & $0.997 \pm 0.074$\\
\nv   & 5  & $  44.0\pm 25.9$ & $    34 \pm   6$ & $   18 \pm  2$ & $0.999 \pm 0.057$\\
\nv   & 6a & $  16.7\pm  7.1$ & $    79 \pm   7$ & $   41 \pm  6$ & $0.997 \pm 0.117$\\
\nv   & 6  & $  46.8\pm  6.8$ & $   121 \pm   6$ & $   12 \pm  1$ & $0.997 \pm 0.117$\\
\nv   & 6b & $   5.2\pm  3.0$ & $   147 \pm   9$ & $   11 \pm  3$ & $0.997 \pm 0.117$\\
\nv   & 7a & $  25.1\pm 16.9$ & $   184 \pm   8$ & $   16 \pm  3$ & $0.704 \pm 0.231$\\
\nv   & 7  & $  22.0\pm 14.8$ & $   222 \pm   8$ & $   16 \pm  3$ & $0.704 \pm 0.231$\\
\siiv & 4  & $   3.5\pm  1.1$ & $   -18 \pm   5$ & $    6 \pm  1$ & $0.946 \pm 0.197$\\
\civ  & 1  & $  71.6\pm 19.6$ & $  -406 \pm   6$ & $   12 \pm  1$ & $0.560 \pm 0.276$\\
\civ  & 1b & $  21.1\pm  5.7$ & $  -365 \pm   5$ & $   27 \pm  8$ & $1.000 \pm 0.110$\\
\civ  & 2  & $ 144.2\pm 24.2$ & $  -320 \pm   6$ & $   14 \pm  1$ & $0.983 \pm 0.078$\\
\civ  & 2b & $ 137.9\pm 23.3$ & $  -292 \pm   6$ & $   30 \pm  5$ & $0.893 \pm 0.090$\\
\civ  & 3  & $  50.5\pm 27.8$ & $  -243 \pm   7$ & $   25 \pm  2$ & $0.963 \pm 0.031$\\
\civ  & 4a & $  67.9\pm  8.5$ & $   -60 \pm   6$ & $   13 \pm  1$ & $1.000 \pm 0.069$\\
\civ  & 4  & $ 233.7\pm 17.9$ & $   -18 \pm   5$ & $   17 \pm  1$ & $0.960 \pm 0.037$\\
\civ  & 5  & $  35.8\pm 45.6$ & $    36 \pm   6$ & $   20 \pm  1$ & $0.916 \pm 0.098$\\
\civ  & 6a & $   0.1\pm  0.7$ & $    88 \pm  68$ & $   12 \pm 56$ & $1.000 \pm 0.045$\\
\civ  & 6  & $  16.1\pm 18.1$ & $   124 \pm   1$ & $   13 \pm  1$ & $1.000 \pm 0.045$\\
\civ  & 6b & $  12.2\pm  7.1$ & $   120 \pm   1$ & $   24 \pm  1$ & $1.000 \pm 0.045$\\
\civ  & 7a & $   1.1\pm  1.2$ & $   171 \pm   6$ & $    6 \pm  3$ & $0.560 \pm 0.276$\\
\civ  & 7  & $   6.1\pm  3.0$ & $   214 \pm   2$ & $    6 \pm  3$ & $0.560 \pm 0.276$\\
\hline                  
\end{tabular}
\\
{The scattered light component in this fit is $0.01 \pm 0.03$.}
\end{table*}

\begin{table*}
	\caption[]{Intrinsic Absorption Features in the 1999 FUSE Spectrum of \mrk509\label{1999FUSE_alines}}
\begin{tabular}{l c c c c c}
\hline\hline       
{Ion} & {Component} & {$\rm N$} & {$\rm v_{sys}$} & {$\rm b$} & {$\rm f_c$} \\
      & { \# } & {($\rm 10^{12}~cm^{-2}$) } & {($\rm km~s^{-1}$)} & {($\rm km~s^{-1}$)} &  \\
\hline
\hi  & 1a  & $  24.4\pm  9.9$ & $  -433 \pm  33$ & $   32 \pm 14$ & $0.991 \pm 0.004$\\
\hi  & 1   & $ 363.3\pm  9.1$ & $  -411 \pm   7$ & $   31 \pm  3$ & $0.991 \pm 0.004$\\
\hi  & 1b  & $   0.3\pm  2.8$ & $  -383 \pm 134$ & $   15 \pm 99$ & $0.991 \pm 0.004$\\
\hi  & 2   & $ 549.4\pm 61.9$ & $  -323 \pm   6$ & $   19 \pm  3$ & $0.978 \pm 0.036$\\
\hi  & 2b  & $  58.1\pm 55.3$ & $  -292 \pm   5$ & $   12 \pm  6$ & $0.826 \pm 0.246$\\
\hi  & 3   & $ 681.5\pm 104.9$ & $  -258 \pm   8$ & $   31 \pm  3$ & $0.869 \pm 0.037$\\
\hi  & 4a  & $1924.1\pm 154.4$ & $   -35 \pm   8$ & $   43 \pm  2$ & $0.939 \pm 0.029$\\
\hi  & 4   & $2594.9\pm 1146.8$ & $   -29 \pm   7$ & $   23 \pm  3$ & $0.902 \pm 0.088$\\
\hi  & 5   & $ 359.5\pm 45.2$ & $    27 \pm   9$ & $   32 \pm  9$ & $0.999 \pm 0.116$\\
\hi  & 6a  & $ 114.3\pm 92.8$ & $    75 \pm  14$ & $   32 \pm  1$ & $0.902 \pm 0.020$\\
\hi  & 6   & $ 520.3\pm 93.7$ & $   115 \pm   6$ & $   32 \pm  1$ & $0.902 \pm 0.020$\\
\hi  & 6b  & $ 168.4\pm  1.7$ & $   129 \pm  13$ & $   32 \pm  1$ & $0.902 \pm 0.020$\\
\hi  & 7a  & $ 154.4\pm 73.9$ & $   174 \pm   8$ & $   19 \pm  4$ & $0.905 \pm 0.154$\\
\hi  & 7   & $ 227.8\pm 118.6$ & $   211 \pm   8$ & $   24 \pm  7$ & $0.969 \pm 0.098$\\
\ovi & 1a      & $   0.7\pm  5.5$ & $  -427 \pm  33$ & $    9 \pm 11$ & $0.697 \pm 0.072$\\
\ovi & 1       & $ 215.0\pm 47.2$ & $  -388 \pm   8$ & $   22 \pm  2$ & $0.697 \pm 0.072$\\
\ovi & 1b      & $ 154.9\pm 39.0$ & $  -360 \pm   5$ & $   22 \pm  2$ & $0.697 \pm 0.072$\\
\ovi & 2       & $ 566.2\pm 118.0$ & $  -318 \pm   5$ & $   19 \pm  3$ & $0.926 \pm 0.076$\\
\ovi & 2b      & $1248.1\pm 395.5$ & $  -290 \pm   5$ & $   17 \pm  2$ & $0.919 \pm 0.037$\\
\ovi & 3       & $ 675.2\pm 157.1$ & $  -247 \pm   5$ & $   28 \pm  3$ & $0.814 \pm 0.048$\\
\ovi & 4a      & $ 804.5\pm 387.2$ & $   -65 \pm   5$ & $   15 \pm  2$ & $0.714 \pm 0.049$\\
\ovi & 4       & $8797.0\pm 4120.3$ & $    -9 \pm   5$ & $   28 \pm  3$ & $1.000 \pm 0.008$\\
\ovi & 5       & $ 683.5\pm 429.3$ & $    47 \pm   5$ & $   24 \pm 22$ & $1.000 \pm 0.032$\\
\ovi & 6a      & $ 518.8\pm 398.5$ & $    87 \pm   5$ & $   15 \pm 21$ & $0.949 \pm 0.160$\\
\ovi & 6       & $3436.1\pm 6985.0$ & $   120 \pm   6$ & $   12 \pm  7$ & $1.000 \pm 0.014$\\
\ovi & 6b      & $ 104.5\pm 234.6$ & $   146 \pm   5$ & $   12 \pm  7$ & $1.000 \pm 0.014$\\
\ovi & 7a      & $ 660.2\pm 176.7$ & $   179 \pm  41$ & $   22 \pm 78$ & $0.834 \pm 0.191$\\
\ovi & 7       & $ 667.4\pm 1105.3$ & $   213 \pm  17$ & $   26 \pm  8$ & $1.000 \pm 0.023$\\
\hline                  
\end{tabular}
\\
{The scattered light component in this fit is $0.04 \pm 0.02$.}
\end{table*}

\begin{table*}
	\caption[]{Intrinsic Absorption Features in the 2000 FUSE Spectrum of \mrk509\label{2000FUSE_alines}}
\begin{tabular}{l c c c c c}
\hline\hline       
{Ion} & {Component} & {$\rm N$} & {$\rm v_{sys}$} & {$\rm b$} & {$\rm f_c$} \\
      & { \# } & {($\rm 10^{12}~cm^{-2}$) } & {($\rm km~s^{-1}$)} & {($\rm km~s^{-1}$)} &  \\
\hline
\hi  & 1a  & $  31.1\pm 41.1$ & $  -425 \pm   8$ & $   29 \pm 25$ & $1.000 \pm 0.071$\\
\hi  & 1   & $ 302.5\pm 41.9$ & $  -402 \pm   5$ & $   28 \pm  2$ & $1.000 \pm 0.071$\\
\hi  & 1b  & $   0.3\pm  2.8$ & $  -371 \pm 118$ & $   15 \pm 19$ & $1.000 \pm 0.071$\\
\hi  & 2   & $ 563.0\pm 167.1$ & $  -318 \pm   5$ & $   24 \pm  5$ & $0.970 \pm 0.078$\\
\hi  & 2b  & $  75.9\pm 69.2$ & $  -281 \pm   8$ & $   16 \pm  7$ & $0.890 \pm 0.072$\\
\hi  & 3   & $ 824.6\pm 51.5$ & $  -250 \pm   5$ & $   28 \pm  1$ & $0.793 \pm 0.061$\\
\hi  & 4a  & $3493.7\pm 12430.4$ & $   -35 \pm  54$ & $   34 \pm 21$ & $0.904 \pm 0.093$\\
\hi  & 4   & $2506.3\pm 9025.3$ & $    -4 \pm  48$ & $   27 \pm 66$ & $0.793 \pm 0.159$\\
\hi  & 5   & $ 327.8\pm 37.1$ & $    47 \pm  53$ & $   25 \pm 20$ & $1.000 \pm 0.027$\\
\hi  & 6a  & $ 175.9\pm 275.9$ & $    89 \pm 107$ & $   31 \pm 30$ & $1.000 \pm 0.027$\\
\hi  & 6   & $ 655.7\pm 1019.0$ & $   123 \pm  12$ & $   31 \pm 30$ & $1.000 \pm 0.027$\\
\hi  & 6b  & $ 138.0\pm 358.2$ & $   166 \pm 138$ & $   31 \pm 30$ & $1.000 \pm 0.027$\\
\hi  & 7a  & $ 365.8\pm 282.3$ & $   185 \pm   6$ & $   17 \pm 44$ & $0.565 \pm 0.394$\\
\hi  & 7   & $ 243.4\pm 81.4$ & $   217 \pm  26$ & $   25 \pm 162$ & $1.000 \pm 0.027$\\
\ovi & 1a      & $  16.5\pm 17.6$ & $  -427 \pm   5$ & $    9 \pm  5$ & $0.697 \pm 0.271$\\
\ovi & 1       & $ 154.9\pm 113.5$ & $  -396 \pm   6$ & $   17 \pm  7$ & $0.697 \pm 0.271$\\
\ovi & 1b      & $ 210.5\pm 123.3$ & $  -363 \pm  15$ & $   17 \pm  7$ & $0.697 \pm 0.271$\\
\ovi & 2       & $ 968.3\pm 187.2$ & $  -312 \pm   5$ & $   18 \pm  4$ & $0.926 \pm 0.025$\\
\ovi & 2b      & $ 125.6\pm 68.6$ & $  -281 \pm   7$ & $   15 \pm  6$ & $0.919 \pm 0.103$\\
\ovi & 3       & $ 917.3\pm 181.2$ & $  -256 \pm   5$ & $   39 \pm  2$ & $0.814 \pm 0.065$\\
\ovi & 4a      & $ 640.6\pm 68.3$ & $   -56 \pm   5$ & $   23 \pm  2$ & $0.924 \pm 0.008$\\
\ovi & 4       & $1293.2\pm 735.3$ & $    -9 \pm   6$ & $   20 \pm  1$ & $1.000 \pm 0.015$\\
\ovi & 5       & $ 569.2\pm 275.9$ & $    36 \pm  10$ & $   20 \pm  1$ & $1.000 \pm 0.052$\\
\ovi & 6a      & $ 324.8\pm 179.7$ & $    64 \pm  13$ & $   20 \pm  1$ & $0.949 \pm 0.082$\\
\ovi & 6       & $ 565.4\pm 46.2$ & $   104 \pm   5$ & $   20 \pm  1$ & $1.000 \pm 0.027$\\
\ovi & 6b      & $ 362.4\pm 59.3$ & $   137 \pm   6$ & $   20 \pm  1$ & $1.000 \pm 0.027$\\
\ovi & 7a      & $ 403.8\pm 187.2$ & $   185 \pm   6$ & $   20 \pm  1$ & $0.834 \pm 0.160$\\
\ovi & 7       & $ 640.3\pm 217.3$ & $   213 \pm   5$ & $   25 \pm  1$ & $1.000 \pm 0.019$\\
\hline                  
\end{tabular}
\\
{The scattered light component in this fit is $0.05 \pm 0.01$.}
\end{table*}

\end{document}